\documentclass[aps,superscriptaddress,reprint,nofootinbib]{revtex4-2}

\usepackage{xcolor}
\usepackage{graphicx}
\usepackage{float}
\usepackage{mwe}
\usepackage{indentfirst}
\usepackage{algorithm}
\usepackage{algpseudocode}
\usepackage{amsfonts, amsmath, amssymb}
\usepackage{mathtools}
\usepackage{xurl}

\graphicspath{ {./figs/} }
\begin{document}
\title{Boosting the effective performance of\\ massively parallel tensor network state algorithms on \\ hybrid CPU-GPU based architectures\\ via non-Abelian symmetries}

\author{Andor Menczer}%
\affiliation{%
Strongly Correlated Systems "Lendület" Research Group,
Wigner Research Centre for Physics, H-1525, Budapest, Hungary
}%
\affiliation{%
Eötvös Loránd University, Budapest, Hungary
}%

\author{Örs Legeza}
\affiliation{%
Strongly Correlated Systems "Lendület" Research Group,
Wigner Research Centre for Physics, H-1525, Budapest, Hungary
}%
\affiliation{
Institute for Advanced Study,Technical University of Munich, Lichtenbergstrasse 2a, 85748 Garching, Germany
}

\date{\today}

\begin{abstract}
	\noindent 
 In this contribution, we present novel algorithmic solutions together with implementation details 
 utilizing non-Abelian symmetries in order to boost the current limits of tensor network state algorithms on high performance computing infrastructure.
 Building on state-of-the-art hardware and software technologies our new mathematical model for parallelization relies on our in-house developed hybrid CPU-multiGPU solution. 
 Scheduling is decentralized, threads are autonomous and inter-thread communications are solely limited to interactions with globally visible lock-free constructs.
 Our custom tailored virtual memory management ensures data is produced with high spatial locality, which together with the use of specific sequences of strided batched matrix operations translates to significantly higher overall throughput. 
 In order to complement higher raw performance with lower IO overhead, an adaptive buffering technique is used to dynamically match the level of data abstraction, at which cache repositories are built and reused, to system resources.
 The non-Abelian symmetry related tensor algebra based on Wigner-Eckhart theorem is fully detached from the conventional tensor network layer, thus massively parallel matrix and tensor operations can be performed without additional overheads.
 Altogether, we have achieved an order of magnitude increase in performance with respect to results reported in arXiv:2305.05581 in terms of computational complexity and at the same time a factor of three to six in the actual performance measured in TFLOPS. 
 Benchmark results are presented on Hilbert space dimensions up to $2.88\times10^{36}$
 obtained via large-scale $SU(2)$ spin adapted density matrix renormalization group simulations on selected strongly correlated molecular systems. 
 These demonstrate the utilization of NVIDIA's highly specialized tensor cores, leading to performance around 110 TFLOPS on a single node supplied with eight NVIDIA A100 devices. In comparison to $U(1)$ implementations with matching accuracy, our solution has an estimated effective performance of 250-500 TFLOPS.
\end{abstract}

\maketitle
%----------------------------------------------------------------------------------------

\section{Introduction}
\label{sec:intro}

There is an ever growing demand for efficient simulation of quantum systems via classical computation\cite{Xu-2023-herculean}. Despite the industry's best effort to create increasingly more powerful hardware~\cite{Nassif-2022,Burd-2022,Munger-2023,Elster-2022,Svedin-2021,Jouppi-2017,Jouppi-2020}, a fundamental limitation emerges: the so-called curse of dimensionality, that is, the computational effort and the dimension of the Hilbert space for systems described by multiparticle Schrödinger equation scale exponentially with the number of constituents~\cite{Feynman-1949a}. Therefore, searching for algorithmic solutions with the potential to reduce the exponential scaling by controlled approximations 
and doing so in a way that modern High-Performance Computing (HPC) ~\cite{anzt2023high} infrastructures are fully taken advantage of, is in the focus of modern quantum physics and chemistry ~\cite{Hager-2004,Stoudenmire-2013,Nemes-2014,Secular-2020,Brabec-2021,Zhai-2021,Gray-2021,Lyakh-2022,Ganahl-2023,Liu-2023,Menczer-2023}.

The density matrix renormalization group (DMRG) method~\cite{White-1992b}, a subclass of tensor network state (TNS) methods, has the potential to fulfill both criteria
as the number of required arithmetic calculations can be reduced by multiple magnitudes. In consequence, the exponential running time collapses into polynomial complexity.~\cite{Schollwock-2005,Noack-2005,Verstraete-2008,Legeza-2008,Chan-2008,Schollwock-2011,Szalay-2015a,Orus-2014,Khoromskaiaab-2015,Baiardi-2020},
In addition, large-scale tensor operations can be substituted with multi-million independent vector and matrix operations, leading to layers of abstractions ranging from low level SIMD\footnote{Single-Instruction Multiple-Data} instructions to high level HPC scheduling.

Quite recently, an efficient hybrid CPU-multiGPU implementation of the DMRG method for complex chemical systems has been introduced~\cite{Menczer-2023} demonstrating a linear scaling in performance with the number of GPU devices, achieving a dramatic decrease in computational time.
Although, the theoretical FP64\footnote{Arithmetic with 64 bit floating-point format} performance ceiling has been reached, the exploitation of non-Abelian symmetries, which is known to have big potential to further boost performance, has not been considered yet.
From technical point of view, utilization of more symmetries is very advantageous as it increases the number of independent vector and matrix operations. However,
an efficient application of non-Abelian symmetries in HPC framework is a highly non-trivial task as it requires a more delicate mathematical framework based on Wigner-Eckhart theorem~\cite{Wigner-1959}. 
In addition, the same accuracy can be achieved via much smaller matrix vector operations, thus scaling properties obtained via a $U(1)$ 
implementation only can be highly modified and efficient simulations often require further algorithmic developments.

It is the main aim of our current work to exploit the benefits of non-Abelian symmetries in TNS algorithms~\cite{Mcculloch-2002,Mcculloch-2007,Toth-2008,Legeza-2008manual,Moca-2012,Sharma-2012,Weichselbaum-2012,Singh-2012,Wouters-2014,Keller-2016,Hubig-2018,Werner-2019,Gunst-2019,Weichselbaum-2020,Werner-2020su2} via hybrid CPU-GPU kernels on HPC infrastructure. We demonstrate that our novel algorithmic solutions can lead to tremendous increase in performance without additional overheads for the multiGPU accelerated simulations. 
We focus on the $SU(2)$ spin adapted version of the DMRG as it can be applied directly to many problems in quantum chemistry, condensed matter physics and nuclear shell models~\cite{Dukelsky-2004,Sharma-2012,Szalay-2015a,Singh-2012}.

The paper is organized as follows. In Sec.~\ref{sec:theory} we present a brief overview of non-Abelian symmetry in the context of tensor network state methods highlighting those technical aspects which are relevant for efficient parallelization. 
In Sec.~\ref{sec:methods} we
introduce methods developed according to various parallelization strategies and novel algorithmic solutions to achieve an efficient hybrid CPU-multiGPU kernel for simulations via non-Abelian symmetries.
In Sec.~\ref{sec:results} we present numerical benchmark results and scaling analysis for selected chemical systems
together with discussions on future possibilities.
Point-by-point conclusions, Sec.~\ref{sec:conclusion}, close our presentation. 
Technical details of the derivation of the Wigner-9j function is collected in an appendix.

\section{Theory}
\label{sec:theory}

In this section we discuss non-Abelian symmetry in the context of tensor network state methods and focus only on those aspects that are relevant for efficient implications on HPC infrastructures. 
Our algorithmic developments 
is presented for the density matrix renormalization group (DMRG) method~\cite{White-1992b} that is a special variant of tensor network state (TNS) algorithms.~\cite{Schollwock-2011,Noack-2005,Chan-2008,Szalay-2015a,Orus-2014,Baiardi-2020}
We focus on a very general form of the Hamiltonian operator, implemented in our code ~\cite{dmrg-budapest},  
that can treat any form of non-local interactions related to
two-particle scattering processes. The corresponding
Hamiltonian can be written in the form
\begin{equation}
\mathcal{H} = \sum_{ij\alpha\beta} T_{ij}^{\alpha\beta} 
                c^\dagger_{i\alpha}c_{j\beta} +
                \sum_{ijkl\alpha\beta\gamma\delta} 
      {V_{ijkl}^{\alpha\beta\gamma\delta}
      c^\dagger_{i\alpha}c^\dagger_{j\beta}c_{k\gamma}c_{l\delta}},
\label{eq:ham}
\end{equation}
where the indices
$\alpha,\beta,\gamma,\delta$ label internal degrees of freedom, like spin or isospin.
The operators $c^\dagger_{i\alpha}$ or $c_{i\alpha}$ usually denote spin ladder or
fermion creation and annihilation operators.
Indices $i,j,k,l$ label, in general, arbitrary modes which allows us to simulate strongly correlated quantum many body problems in various fields of disciplines, like condensed matter physics, nuclear structure theory or quantum chemistry even in the relativistic domain
~\cite{White-1992b,Xiang-1996,White-1999,Knecht-2014,Dukelsky-2004,Legeza-2015,Legeza-2018a,Shapir-2019,Barcza-2020}.

The related eigenvalue problem is obtained by the iterative diagonalization of the so-called effective quantum many body Hamiltonian ~\cite{Schollwock-2011} via the L\'anczos or Davidson algorithms corresponding usually to 85-90\% of the total execution time.
In the TNS/DMRG methods renormalized block operators are formed via the course of the network contraction procedure
which is responsible for another 5-10\% of the total execution time. 
The underlying TNS/DMRG tensor and matrix algebra
can be organized into several million of independent operations (tasks) and  
performed in parallel 
according to the so-called quantum number decomposed representations.
The size of the full matrices, denoted as DMRG bond dimension, $D$, determines the accuracy of the calculations and at the same time the required
computational complexity. The overall scaling of the DMRG for models described by Eq.~(\ref{eq:ham}) is $D^3N^4$ where $N$ stands for the number of modes, i.e., for the system size. The memory requirement is proportional to $D^2N^2$.
The details of the algorithm can be found in various review articles \cite{Schollwock-2005,Schollwock-2011,Noack-2005,Szalay-2015a,Chan-2008,Orus-2014,Baiardi-2020}.

In our implementation, the non-Abelian symmetry related complex mathematical models, that are computationally lightweight as no actual data transformation occurs,
are handled via MATLAB\footnote{MATLAB is a high level interpreter based language for modeling complex systems.}. 
In addition, since GPU devices are handled at a lower abstraction level, all mathematical aspects of the DMRG algorithm is considered on the host (CPU) side.
In contrast to this, the entire number crunching computational part is implemented in native C++, and GPU devices are used only for executing basic algebraic operations in large batches.
In the followings, we discuss only those algorithmic developments which are relevant for boosting the performance via non-Abelian symmetry, while for further details on our hybrid CPU-multiGPU kernel we guide the readers to Ref.~\cite{Menczer-2023}.

\subsection{Hierarchy of tasks}
\label{ssec:dmrg-integration}

%%---operators, hamtable
In DMRG the modes of a network are partitioned into subsystems (blocks) and for efficient treatment of long range interactions, pre-contracted operators are formed to reduce the computational complexity from $N^4$ to $N^2$.~\cite{Xiang-1996,White-1999} The Hamiltonian in Eq.~(\ref{eq:ham}) is construed from these operators using the matrix product operator (MPO) representation~\cite{Schollwock-2005,Noack-2005,Schollwock-2011,Szalay-2015a,Keller-2015,Chan-2016}.
In the two-site DMRG topology the related matrix vector multiplications in the iterative diagonalization via the Davidson or L\'anczos algorithm is obtained from an accumulated sum of four matrix multiplications along four distinct dimensions of a four-dimensional tensor, i.e.,
\begin{equation}
F \coloneqq F + \sum_{i=1}^q \alpha_i A_i B_i C_i D_i E\,,
\label{eq:hammultpsi}
\end{equation}
where $q$ stands for the number of independent operator combinations in the MPO,
$\alpha_i$ is a precalculated constant, $A_i,B_i,C_i$ and $D_i$ label operators of the four subsystems, and
$E$ and $F$ are the 4-index tensor representations of the quantum many body wave function. 
In our implementation, each of such operator combination is stored in a table (operator-table).

%%---sectors, tasktable
In the TNS/DMRG algorithm, the matrices and tensors are decomposed into smaller components (sectors) based on quantum numbers, i.e., a full matrix is stored according to row-column quantum number sector pairs (sector-table) and the corresponding dense matrix ~\cite{Nemes-2014,Szalay-2015a,Brabec-2021,Menczer-2023}. 
All operations of the underlying DMRG linear algebra is developed via such sector representation by storing sector combinations in lookup tables (sector-task-tables). 
For example, for a given $i$ in Eq.~(\ref{eq:hammultpsi}) 
the algebra is decomposed into a series of elementary
tensor operations, where sector components of the matrices
and tensors are address by six integer pointers, $p^{\rm K}_n$ with $K\in\{A,B,C.D,E,F\}$, stored
in rows of the sector-task-table having as many rows, $\chi_i$, as the given operation would decompose into, i.e., 
\begin{equation}
\begin{aligned}
F\{p^{\rm F}_n\} \coloneqq & F\{p^{\rm F}_n\} + \\
&\alpha_iA_i\{p^{\rm A}_n\} 
B_i\{p^{\rm B}_n\}
C_i\{p^{\rm C}_n\}
D_i\{p^{\rm D}_n\}
E\{p^{\rm E}_n\}\,,
\end{aligned}
\label{eq:hammultpsi-sector}
\end{equation}
with $n=1,\ldots, \chi_i$.
%

%%---supersize, pos
For long range interactions, the contracted block operators are position dependent, thus in our tensor library ~\cite{tenlibol}, the sector dependent dense matrices of the same operator type have a third index as well, labeling their positions in the given subsystem.
These form contiguous segments in memory, while operations 
often access only subsets of them which requires clever on-the-fly regrouping in order to achieve an efficient and  highly parallel execution of the given algebra.

Therefore, in our implementation, each operator combination of an algebra is stored in an operator-table and each operator combination is supplied with the corresponding sector-task-table. As a result, there are three main loops that must be executed: loop over the rows of the operator-table, for each row a loop over the rows of the corresponding sector-task-table and finally a loop over the position index within the subsystem blocks. The product of these provides the number of independent operations (tasks) that must altogether be executed to perform a given algebra~\cite{Nemes-2014}. 
Since the sector sizes and position ranges are available a prior of the execution of a given algebra the computational complexity can be precalculated and various dynamic scheduling algorithms can be developed optimizing overheads.
As discussed in Ref.~\cite{Menczer-2023}, there are multiple methods for organizing the underlying DMRG algebra into independent tasks, with each varying in asymptotic space, IO time and compute time complexities. 
In Sec.~\ref{sec:methods}
we present the importance of task hierarchy, i.e., how to group subsets of tasks together at different abstraction levels in order to maximize performance and minimize IO overhead. 
This is mandatory for GPU accelerated
DMRG calculations on model Hamiltionians with long range interactions as they require large amount of data stored in RAM that far exceeds current GPU memory sizes~\cite{Nemes-2014,Menczer-2023}.

\subsection{Utilization of non-Abelian symmetry}
\label{ssec:NA}

In case of $SU(2)$ symmetry the matrix and tensor algebra in TNS/DMRG can be performed just like for $U(1)$ symmetry except that the Hamiltonian is reformulated in terms of so-called reduced operators.~\cite{Mcculloch-2002,Mcculloch-2007,Toth-2008} These operators have smaller dimensions compared to the original operators defined for $U(1)$ symmetry. For example, for a spin-1/2 fermion model
given by Eq.~(\ref{eq:ham}) in case of $SU(2)$ spin symmetry out of the four local basis expressed in occupation numbers together with spin degrees of freedom, $|0\rangle, |\downarrow\rangle, |\uparrow\rangle |\uparrow \downarrow\rangle$ only three remains as the $|\downarrow\rangle$ and $|\uparrow\rangle$ are connected via the $SU(2)$ spin symmetry. In certain special circumstances, when $SU(2)$ charge symmetry is also applied, the dimension of the local Hilbert space shrinks to two as  
bases $|0\rangle$ and $|\uparrow\downarrow\rangle$ also transform to each other.~\cite{Toth-2008,Legeza-2008manual} Accordingly, reduced operators with dimension three or two are formed via the Wigner-Eckhart theorem ~\cite{Wigner-1959,Cornwell-1997} and after the related rescaling of the couplings constants in Eq.~(\ref{eq:ham}) the non-Abelian version of the Hamiltionan is achieved. ~\cite{Mcculloch-2002a,Toth-2008,Sharma-2012,Wouters-2014,Keller-2016}
The tensor and matrix algebra, however, requires further corrections as reduced operators also posses quantum numbers, that are invariant under transformation of the given non-Abelian symmetry.~\cite{Wigner-1959,Cornwell-1997} These scalar multiplication factors are determined by the spin of the bra and ket  states, by the spin of reduced operators appearing in the given operation and by the corresponding Clebsch-Gordan coefficients.\cite{Cornwell-1997}
Here we summarize only those formulas that are used in the context of the current work using a Matlab implementation of the so called Wigner-9j and Wigner-6j formalism ~\cite{Varshoalovich-1988,libsu2,Weisstein-0000} while further details can be found in the original publications.~\cite{Mcculloch-2002,Toth-2008,Legeza-2008manual,Moca-2012,Sharma-2012,Weichselbaum-2012,Singh-2012,Wouters-2014,Keller-2016,Hubig-2018,Gunst-2019,Werner-2019,Weichselbaum-2020,Werner-2020su2}

More precisely, when matrix elements are calculated for a tensor product operation between bra and ket states their total spin $j$ and the internal degrees of freedom, $-j\leq m\leq j$, must also be taken into account. In addition, tensor product of basis
states with $j_1$ and $j_2$ quantum numbers can generate matrix elements with quantum number in the range of $|j_1-j_2|\leq j\leq j_1+j_2$. Similar arguments hold for operators with spin $k_1$ and $k_2$, i.e., the resulting operator can have spin $|k_1-k_2|\leq k\leq k_1+k_2$. To fix notations, we consider a tensor product operation where operators have spin $k_i$ and states are labeled as $|j_i, m_i\rangle$ with $i\in\{1,2\}$, i.e.
\begin{equation}
\begin{aligned}
k_1,k_2 & \rightarrow k\\   
\langle j_1,m_1| \otimes \langle j_2,m_2| & \rightarrow  \langle j,m| \\        
|j^\prime_1,m^\prime_1\rangle \otimes |j^\prime_2,m^\prime_2\rangle & \rightarrow  |j^\prime,m^\prime\rangle\,.  
\end{aligned}
\end{equation}
The corresponding scalar factor that modifies the conventional tensor product operation developed for $U(1)$ symmetry, $\Tilde{C}$, can be determined for non-Abelian $SU(2)$ symmetry
as
\begin{equation}
\begin{aligned}
    \tilde{C} = & \sqrt{(2j_1^\prime+1)(2j_2^\prime+1)(2j+1)(2k+1)}\times \\
    & W_{\rm 9j}(j_1,j_2,j,k_1,k_2,k,j_1^\prime,j_2^\prime,j^\prime)\,, 
\end{aligned}
\label{eq:a}
\end{equation}
where the detailed derivation of $W_{\rm 9j}$ is collected in the Appendix (see Eqs.~(\ref{eq:w9})-(\ref{eq:del})). %~\ref{sec:appendix}.
As a result of the Wigner-Eckhart theorem, the underlying TNS/DMRG matrix and tensor algebra can be reformulated according to 
the non-Abelian quantum numbers. Therefore, instead of keeping renormalized states corresponding to $U(1)$ symmetries, they are selected according to quantum numbers of multiplets~\cite{Mcculloch-2007,Toth-2008}. 
Usually a factor of two to four reduction in bond dimension can be achieved for the same accuracy which can lead easily to an order of magnitude speedup as will also be demonstrated in Sec.~\ref{sec:results}.

\subsection{Required modification for non-Abelian symmetry and efficient implementation on HPC infrastructure}

For models in which the local Hilbert spaces of the two-intermediate modes are decomposed into one dimensional sectors based on quantum numbers
-- like in the model systems studied in this paper --, 
the complexity of Eq.~(\ref{eq:hammultpsi})
can be reduced to a series of matrix multiplications
by preprocessing the corresponding sector-task-table based on Eq.~(\ref{eq:hammultpsi-sector}), and storing the product of the sector contributions of $B_i\{p^{\rm B}_n\}$, $C_i\{p^{\rm C}_n\}$ and $\alpha_i$ in an additional table (sector-value-table) ~\cite{Menczer-2023}.
This is advantageous when operations given by Eqs.~(\ref{eq:hammultpsi}) are executed several times, for example, in case of the diagonalization procedure.
Using the formalism outlined by Eqs.~(\ref{eq:w9}-\ref{eq:del}) the Clebsch-Gordan layer can be separated from the
MPS layer (see also Refs.~\cite{Werner-2019,Weichselbaum-2020}) leading to an additional scalar multiplication only for both the matrix and tensor product operations.
Preprocessing a given sector-task-table not only provides pointer structures addressing proper sector components of the  
operators $A_i, B_i, C_i, D_i, E$ and $F$ in Eq.~(\ref{eq:hammultpsi}), but also gives the corresponding quantum numbers. These can be used immediately to precalculate the corresponding scalar factors derived in Sec.~\ref{ssec:NA} to include contributions of the    
Clebsch-Gordan layer and modify the related entries in the sector-value-table.

For example, the tensor product operation, $A\otimes B$, in case of $SU(2)$ spin symmetry is performed like in case of $U(1)$ symmetry, i.e., tensor products from sector components of $A$ and $B$ are formed by processing the corresponding sector-task-table, but each result is multiplied with an additional scalar
given by Eq.~(\ref{eq:a}).
The scalar corrections, $\Tilde{C}$, for tensor and matrix
operations in 
Eq.~(\ref{eq:hammultpsi-sector}) can be determined similarly.
However, the $W_{9j}$ in Eq.~(\ref{eq:a}) is a complicated function and a computationally demanding object, thus for an efficient implementation it is
mandatory to precalculate and store its matrix elements.
Since the size of such an object can be very large, even exceeding the available RAM size for a given problem,
it is split up into three smaller components, that we
label as $W_{9j}^{\rm (l)}$, $W_{9j}^{\rm (r)}$ and $W_{9j}^{\rm (h)}$.
Assuming blocks with highest spin value $S$, the intermediate site with $s$, and operators with spin $S_{\rm O}$ all components of 
the $W_{9j}^{\rm (l)}$ tensor are precalculated for index ranges 
$j_1\in\{0,\frac{1}{2},1,\ldots,S\}$,
$j_2\in\{0,\frac{1}{2},1,\ldots,s\}$,
$j_1^\prime\in\{0,\frac{1}{2},1,\ldots,S\}$, 
$j_2^\prime\in\{0,\frac{1}{2},1,\ldots,s\}$,
$k_1\in\{0,\frac{1}{2},1,\ldots,S_{\rm O}\}$,
$k_2\in\{0,\frac{1}{2},1,\ldots,S_{\rm O}\}$,
$j\in \{|j_1-j_2|,\ldots,j_1+j_2\}$,
$j^\prime\in \{|j_1^\prime-j_2^\prime|,\ldots,j_1^\prime+j_2^\prime\}$,
and $k\in \{|k_1-k_2|,\ldots,k_1+k_2^\prime\}$ and stored.
Practically, for each $j_1,j_2,k_1,k_2,j_1^\prime,j_2^\prime$
index-tuple a 3-dimensional matrix is stored, i.e. it has the size of
\begin{equation}
W_{9j}^{\rm (l)}:[2S+1,2s+1,x,2S_{\rm O}+1,2S_{\rm O}+1,x_k,2S+1,2s+1,x^\prime],
\end{equation}
where $x,x_k,x^\prime$ is determined by the other six indices.
The $W_{9j}^{\rm (l)}$ tensor can be used as a lookup table for the renormalization step when tensor product of the block and the site is formed.
Similarly, when the tensor product of the site and block is
formed the non-Abelian symmetry related scalar factors are taken from $W_{9j}^{\rm (r)}$ that is precalulated and stored in the form:
\begin{equation}
W_{9j}^{\rm (r)}: [2s+1,2S+1,x,2S_{\rm O}+1,2S_{\rm O}+1,x,2s+1,2S+1,x].
\end{equation}
The scalar factors for each operation in Eq.~(\ref{eq:hammultpsi-sector}) is modified with corresponding components of
$W_{9j}^{\rm (l)}$, $W_{9j}^{\rm (r)}$, and $W_{9j}^{\rm (h)}$.
Here, $W_{9j}^{\rm (h)}$ is an object with size
\begin{equation}
W_{9j}^{\rm (h)}: [2S+1,2S+1,1,2S_{\rm O}+1,2S_{\rm O}+1,1,2S+1,2S+1,1]\,,
\end{equation}
where $j=j^\prime=S_{\rm TG}$ as the spin of the target state is fixed, and $k=0$ as the Hamilton is a spin zero object. 
The precalculation of $W_{9j}^{\rm (l)}$ and $W_{9j}^{\rm (r)}$ is computationally demanding while $W_{9j}^{\rm (h)}$ is a small object and calculated for a given target state at the beginning of the DMRG calculation. If data set stored in $W_{9j}^{\rm (l)}$ or $W_{9j}^{\rm (r)}$ is overindexed via the course of the calculations the required tensor element is calculated and the W tensor is updated. Choosing, for example, $S=19/2$, $s=2$, $S_{\rm O}=2$ we have not experienced such scenario for problems studied in this work.

Having $W_{9j}^{\rm (l)}$, $W_{9j}^{\rm (r)}$ and $W_{9j}^{\rm (h)}$ in hand, in a given DMRG iteration step for each entry of the operator-table the supplied sector-task-table is prepocessed and sector and operator dependent non-Abelian scalar correction factors are determined and stored in the sector-value-table. 
This means that actual execution of the independent tasks given by the tables follows only after such initialization procedure is completed, which makes our implementation ideal for HPC infrastructures via dynamic scheduling protocols even in case of non-Abelian symmetries.
Although, in our discussion we have restricted ourselves to a single non-Abelian symmetry and only for the $SU(2)$ spin symmetry, the presented algorithmic solution works in general~\cite{Weichselbaum-2020,Werner-2020su3} once the proper mathematical treatment of the given symmetry is implemented.

\section{Algorithmic solutions}
\label{sec:methods}

In this section, we present details of the various algorithmic solutions of our new model of parallelization that are key to boost the performance for non-Abelian symmetries.

\subsection{Partial Strided Batched Matrix Multiplication for Summation}

Unlike prior designs, we no longer treat the multiplications of matrix arrays as a sequence of GEneric Matrix Multiplications (GEMMs) or as a singular Strided Batched GEMM (SBGEMM) operation, but rather as a sequence of SBGEMMs. This enables us to exert more finesse over the grouping of same-sized matrices. In particular, when a matrix array is deemed unfit for concurrent multiplication, it might still be possible to create smaller, SBGEMM compatible batches within the array.

By decomposing the matrix arrays into such subgroups and dealing with each of these with a single SBGEMM, the overall multiplication of the entire matrix array will yield significantly higher parallelisation compared to sequential GEMMs as long as the matrices residing in these subgroups can be numerous. In other words, instead of interpreting matrix arrays as either Single-Instruction Single-Data (SIMD) batches that are compatible with single SGEMM operations or Multiple-Instruction Multiple-Data (MIMD) batches that are compatible with sequences of GEMMs, we interpret all matrix arrays as MIMD batches of SIMD sub-batches compatible with sequences of SBGEMMs.

This is not an alternative solution, but a more general approach, because in special cases --- by restricting the size of the MIMD batch or all SIMD sub-batches to a single element --- we end up with either a single SBGEMM operation or a sequence of one matrix long SBGEMM operations, which are effectively just ordinary GEMMs.

\begin{figure}
    \centering
    \includegraphics[width=0.48\textwidth]{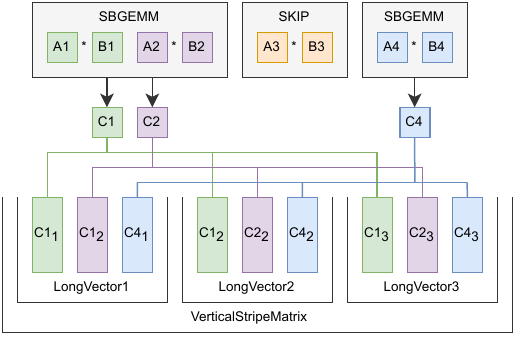}
    \caption{Example for Partial SBMM4S. Input matrices are stored consecutively in memory, however for algorithmic reasons some multiplications can be skipped. Because of this, the multiplication has to be broken up into multiple SBGEMM operations. The leading dimensions and stride values are set in a way that the vectors of the result matrices became interleaved. Since the vectors from different result matrices are next to each other in memory, each vector group can be reinterpreted as a single long vector. Similarly, the sequence of such vectors can be viewed as an either horizontally or vertically very long stripe-like matrix. Using such a matrix as the left or right operand for an ordinary matrix multiplication will work the same way as it does for regular SBMM4S: the entire operation takes only a single kernel call with no additional reduction step needed. Please note that the construction of long vectors and the stripe matrix are handled by the SBGEMM operations, thus they pose no additional overhead.}
    \label{fig:sbgemm}
\end{figure}

Based on whether matrices are stored in column-major or row-major format, consecutively stored matrices can be re-interpreted as a single matrix with a magnitude higher column or row count. Such a matrix will take shape as either a horizontally or vertically very long stripe-like structure. In order to make standard matrix multiplication possible, a horizontally stacked stripe needs to be paired with a vertically stacked one made from the same number of matrices. In both column-major and row-major formats, one of these two operands are naturally available as long as the building blocks (the matrices) are stored consecutively in memory. The other operand can be created through proper usage of leading dimensions and stride value of SBGEMM operations determining memory offsets
as explained in detail in Ref.~\cite{Menczer-2023} including pseudocodes as well.

During matrix multiplication, the dimensions of the resulting matrix is defined by the rows of the left operand and columns of the right, and as such, the result from the multiplication of two compatible stripe matrices have the same dimensions as the multiplicative result of its building blocks. Multiplying $m$ by $k$ sized matrices with $k$ by $n$ sized ones a total of $c$ times is equivalent in terms of computational complexity to multiplying a single $m$ by $k*c$ horizontal stripe matrix with a $k*c$ by $n$ vertical stripe matrix. However, in the latter case the results from elemental multiplications are added to the same output matrix, thus no additional reduction step is necessary.

Please note that while sequentially executed GEMM operations can be directed into the same output matrix, making the output a sum of all prior multiplications, batched GEMM operations require the result matrices to not overlap. In the end, with traditional GEMMs or SBGEMM we either give up parallelism or create additional overhead in the form of having to reduce multiple results into one as a last step. By utilizing stripe matrices multiple matrix multiplications can be merged into a single kernel without creating an unnecessary reduction step.

Stripe matrices can be created independently. Furthermore, as the batch of batched matrix operations create the same stripe matrix, regardless of how the matrices are organized into SBGEMM groups, different batches corresponding to different stripe matrices can be grouped independently. This level of freedom holds true for the final GEMM operations as well. In other words, merging two stripe matrices through multiplication can be done via a single GEMM operation or can be broken up into multiple GEMM operations in case the naturally occurring stripe matrix is segmented (not contiguous in memory).

\subsection{Contractor Threads}

In theory, perfect software scalability is attainable via static scheduling if the executable tasks are independent, same-sized and the program can produce an arbitrary number of these tasks. With such constraints, the real world usage of static schedulers are fairly limited. To overcome the problem of suboptimal load balancing, tasks are often assigned to processors during run-time. Doing so, however, requires dynamic scheduling, which is in itself a program that has to run alongside the computation. This can become a problem when tasks are small enough and defining an optimal distribution of tasks among workers are nontrivial. 

By increasing the number of workers, we expect the total running time to decrease. However, scheduling more and more workers in less and less time becomes increasingly more difficult. Eventually, the scheduling duties will outgrow the computational tasks in complexity and become a severe bottleneck for the application. Unless, that is, the computational burden of the scheduler is also distributed among the working threads, along with the original tasks. Thus, both the total overhead and scheduling speed arising from the parallelization can grow proportionally with the level of parallelization. This way not only the computation, but the parallelization model itself becomes scalable with the number of worker threads.

Standard inter-thread communication protocols such as event or message based models, shared structures relying on mutual exclusion, locking mechanisms and other heavy parallel constructs can slow down or even halt the execution of certain threads while paving the way for just one. In other words, introducing a new workers to the system might do more harm than good for overall performance, if the complexity of thread management rises faster than the computational gains from higher task throughput.

To battle the above mentioned problems, we designed a system in which all threads are self scheduled and managed. Meaning the execution of a single task on a thread must not rely on any activity that might block other threads from doing the same. Furthermore, idle threads must chose available tasks for themselves from a locking-free database, based on information gathered by non-blocking constant time complexity functions.

By following the aforementioned paradigms, the additional overhead from each newly introduced worker can be contained within the new threads themselves, with no possibility of bleeding into existing threads. With the lack of central management and workers having only a loose connection through a catalogue of tasks from which they are free to choose from, this design inherently supports a heterogeneous mixture of workers such as pinned threads for different types of CPUs, GPUs and FPGAs.

\subsection{Contract Book}

Idle workers are obliged to sign up for new tasks. In order to make this endeavour go smoothly, threads can browse a catalogue of available tasks called the Contract Book. This is an object shared among all threads and mostly constitutes of read-only metadata describing the contained tasks. This way workers are not only given new tasks, but tasks that their own scheduling logic finds desirable. Once the preferred task is found, the option can be locked in and made visible for other workers by flipping an atomic boolean based flag, which indicates a task's availability.

Tasks are organized into larger groups based on their dependency on particular data. Accessing tasks through groups makes task selection faster and can help certain workers --- especially those assigned to devices with limited on-board memory --- to chew through tasks in an order that guarantees optimal memory management with minimal IO operations between host and device. Other workers might be less lenient to stick to a specific group of tasks. They might go for the largest or smallest of tasks in each group, clear out leftovers or look for other special cases.

In order for workers to be able to choose the most suitable group of tasks, the groups themselves carry metadata akin to contained tasks. In fact, we can view groups as bigger tasks composed by smaller tasks. This also means entire groups can be chosen by a worker via flipping the group's availability flag. Doing so locks out other workers from choosing new smaller tasks within the group, however, ongoing processes are unaffected and their corresponding smaller tasks will be unavailable for the worker that flipped the group flag.

Given the Contract Book's locking-free nature, workers are never blocked during task selection. Signing up for a certain job might fail however. Failing to flip an availability flag can only mean another worker has taken the task in the meantime. Thankfully this does not result in a slowdown, because the worker simply skips all operations related to the chosen task and immediately starts looking for the next one. Hence, while introducing new workers into the system might increase tasks being snatched from already present workers, this only results in a speed up and earlier termination of the threads that failed to lock in their selection.

\subsection{Hash Based Semi-Dynamic Scheduling}

So far we were focusing on uniting workers into a loosely connected system with low overhead and locking-free task management devoid of any kind of performance hindering interference between threads. What is left to discuss is how workers can effectively utilize their given autonomy for maximal performance gains. In this section we will primarily focus on GPU based systems, but the presented framework offers enough flexibility to effectively utilize other hardware configurations as well. 

On the CPU side of things, the workflow can be made very simple, because in terms of raw theoretical performance even the most powerful CPUs tend to be dwarfed by modern NVIDIA accelerators. There is, however, an addressable problem for CPU based task execution: littering the GPU pipeline with minuscule tasks would lead to significant performance drops. For this reason the CPU based workers are tasked to execute all calculations below a certain threshold of computational complexity. This cutoff point is essentially where the GPU becomes efficient at calculations, while also being small enough for the CPU to finish before the GPU does --- and by doing so avoids CPU bottlenecking. Therefore the optimal value is system dependant, so while that threshold is unknown at compile time, it is always known at runtime, meaning the target complexity for CPU cutoff is known by all workers at all times and can be part of each thread's built-in static scheduler.

In the current context, by static scheduling we mean scheduling related information that is imprinted into the self-scheduler of every worker object by the object's own constructor. This information must not change during the object's life cycle, thus static elements of the schedulers are essentially free of overheads and part of how workers naturally behave without requiring any feedback from the system in which they operate in.

Since a constantly changing scheduling policy would require the worker to track changes or capture meaningful events from the environment, it becomes clear that covering as much scheduling duty as possible by the worker's much more basic static scheduler is desirable. Extending static coverage beyond basic functionality is possible by creating heuristics that try to guess the task with the highest likelihood of being chosen by a dynamic scheduler --- that is aware of other workers and their current situation.

\begin{figure}
    \centering
    \includegraphics[width=0.48\textwidth]{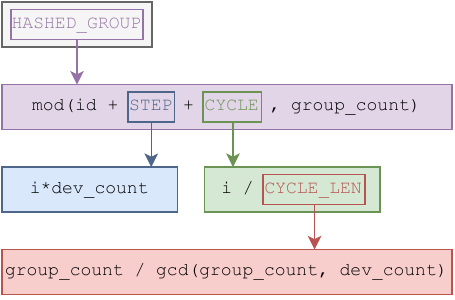}
    \caption{Hash function used for choosing the next group of tasks. Ideal for GPU based workers. Workers initially start with the groups based on their unique identification (id). After the current group has been exhausted, the selection is shifted by the number of GPU based workers (STEP). Once this initial subset of available groups is done, the worker re-iterates over the possible groups and will pick one that belongs to an another worker's initial subset. This is handled by the modulo function (\text{where group\_count} is the total number of groups). However, re-iterating over and over this way does not necessarily make all groups reachable by the worker. We will only generate all possible group indexes if the number of groups and the number of GPU based workers are relative primes. To combat this issue we pre-calculate the length of the cyclical subset of groups (\text{CYCLE\_LEN}). Once this value is known, the  variable \text{CYCLE} will store how many loops we have done so far and we will shift our selection --- in addition to shifting from STEP --- by the same amount as the value itself. By doing so we ensure different groups are selected in every cycle.}
    \label{fig:hash}
\end{figure}

The key to writing such heuristics is predictability. All types of workers are known at compile time and static schedulers are also aware of the CPU/GPU configuration of the current system. By understanding the strategies of all possible worker types and knowing the optimal number of each type for the current hardware, an optimal strategy can be pre-calculated.

Sticking to this global strategy is effortless as individual workers are oblivious to such strategies and simply pick tasks one by one based on a constant time complexity hash function that seemingly just happens to comply with this global strategy when viewing multiple workers as a team. In other words, workers schedule themselves on an individual level, however, the difference in the order in which workers iterate through tasks leads to spontaneously synergetic behaviour between team members. 

To summarize our findings, a good task selecting hash function has to follow the following rules:
\begin{itemize}
    \item Constant time complexity: hashing should be near instantaneous and must not slow down with larger system and model sizes.
    \item Individual level: always selects one of the most desirable tasks for the worker. Solving a sequence of desirable tasks must yield high performance. For GPU based workers, this mainly boils down to batched matrix operations where the computational complexity is high enough to saturate the device, while at the same time also featuring reoccurring data in subsequent instructions in a way that data already present in device memory can be used multiple times. By doing so, the compute to IO kernel ratio can be increased. 
    \item Team level: taking globally visible tasks is the only possible interaction between workers. A good team player will get rid off tasks undesirable by others and will not ruin a worker's streak of selecting highly desirable tasks by taking one of the tasks for itself.
\end{itemize}

CPU based workers find all small tasks equally desirable, hence any order will suffice. This construction is essentially just a parallel \textit{for loop} with dynamic scheduling, meaning tasks are not pre-distributed, but instead, workers come back again and again to take the next task not yet taken by others.

With smaller tasks out of the picture, all remaining tasks are suitable for GPU accelerators. Their order of execution, however, can make or brake the effectiveness of our strategy. Assigning tasks from different groups in the Contract Book to each worker ensures assigned tasks are undesirable for others, due to different groups requiring different sets of matrices to be loaded in memory. Each GPU moves to a new group after the previous one is exhausted.

Eventually, the number of untouched groups will fall below that of the GPU based workers. This is where hashing truly becomes useful over simpler constructs such as a parallel for cycle. With hashing we can deliberately cause collisions, meaning multiple workers --- despite having different unique IDs --- will be assigned to the same group of tasks. By doing so the remaining groups will be finished at an enhanced speed. This way our model can dynamically scale between highest task throughput (focusing on maximum number of solved tasks per given time) and lowest group execution time (focusing on minimizing the time it takes to finish a given group of tasks).

In the initial phase of the computation we can reap the benefits of having statically scheduled disjoint sets of independent jobs, while still having the late-stage option to iron out scheduling imperfections and make runtime optimizations as our complex multiGPU computation progresses. In Fig.~\ref{fig:hash} we show an adequate hashing function, which we also chose for our implementation.

\begin{figure}
    \centering
    \includegraphics[width=0.40\textwidth]{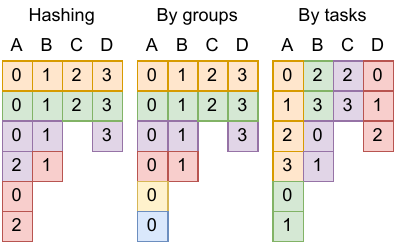}
    \caption{Solving four groups of tasks (A, B, C and D) with different schedulings. Each task is represented with a square, while the number inside is the unique identifier of the worker that solved the task. For simplicity's sake, each task takes the exact same time to execute. This way the colors can show which tasks were solved in parallel. Assigning different groups to different workers (scheduling by groups) can result in unwanted idle time due to inconsistencies between the sizes of groups. Ignoring groups and treating the Contract Book as a long queue of tasks (scheduling by tasks) results in high IO overhead due to many devices taking tasks from the same groups. Hashing on the other hand assigns different groups to different workers whenever possible, but at the same time allows multiple devices to work within the same group if necessary.}
    \label{fig:task_order}
\end{figure}

\subsection{Dedicated IO Streams and Memory Mapping}

Instead of relying on NVIDIA's own virtual memory tables through regular use of \textit{cudaMalloc} and \textit{cudaFree}, we allocate memory only once and use our own model to map memory and enable H2D data transfers. Our solution has been already discussed in Ref.~\cite{Menczer-2023}.

Our newest implementation, utilized in the current work, augments performance by executing H2D kernels on desynchronized IO streams during computations. Encapsulation has also been improved as memory tables are now private and all memory management related optimizations are now handled internally, without the compute algorithm having any visibility on how the requested data was produced on the particular device.

\subsection{Dynamic Memory Buffer Utilization}

Locality of reference can be exploited at multiple levels of our computation. First of all, basic arithmetic operations belonging to the same task have exceptionally high spatial data locality due to position dependent index ranges and TTcache's ability to place data in the order of execution without any gaps in-between elements~\cite{Menczer-2023}. Secondly, tasks relying on the same set of matrices are grouped together. Finally, during Davidson algorithm, the entire set of groups of tasks might be reused at a later time.

IO operations can be reduced dramatically by increasing the length of GPU buffers to accommodate whole sets of data meant for higher levels of computation. Fixing our model at a level we find reasonable would result in polynomial space complexity. However there is no benefit to having unused memory, just as we have no use for computation that would exceed our current hardware's memory limits. Thus, we expect the optimal approach to use all available memory as buffer and dynamically scale between the levels. This way --- as matrix sizes grow --- the algorithm slowly transforms from an IO call optimized, but memory hungry approach to a space optimized, albeit slower variation.

Helped by the fact that streamed data can be placed into the buffer sequentially, it is effectively effortless to store immutable data from left to right and mutable from right to left. The space in-between can act as a temporary smaller buffer for single-use matrices produced locally on the device. Whenever the two sides get too close, the read-only matrices on the left side can be deregistered and eventually overwritten by newer data. This violates the rule of always keeping ancestral data for TTcache nodes, however since the dataset is never modified by the device, it can be recovered from host memory at any point during the worker's life cycle.

Reconstructing previously deleted data from host memory comes at an IO penalty, however, it only happens when available memory is insufficient. Thus, the IO overhead from re-downloads will be based on the level at which deletions are frequent. Multiple deletion without reaching a leaf node indicates that the complete path from the root does not fit into device memory and --- while our computation might suffer from high IO overhead due to low level deletes --- at the very least we made a computation possible that otherwise would have caused memory overflow.

Also, by omitting deletes based on the traversal of the data dependency tree and solely relying on the preemptive deregistration of immutable data as described above, it is possible to store multiple root to leaf paths in memory at the same time. This further reduces IO calls as different paths are not required to be disjoint sets of data. When memory is sufficient, even the entire data dependency tree can be held in device memory at all times, thus reducing IO calls for subsequent runs to zero.

%----------------------------------------------------------------------------------------

\section{Numerical results}
\label{sec:results}

In this section, we present benchmark results obtained via  large-scale $SU(2)$ spin adapted density matrix renormalization group simulations on selected strongly correlated molecular systems.
The F$_2$ molecule in CAS(18,18) orbital space was chosen as it corresponds to current limit of exact diagonalization, the nitrogen dimer is often used as a benchmark system as it has a notorious character in chemical bond formation~\cite{Chan-2004b}, and the FeMoco cluster which is in the focus of theoretical chemistry due to its important role in nitrogen fixation -- i.e., reduction of nitrogen (N$_2$) to ammonia (NH$_3$)~\cite{Hoffman-2014} --
which is essential for the biosynthesis of nucleotides like DNA underlying all life forms on earth. 
In addition, it also serves as a large benchmark system studied in various previous works~\cite{Reiher-2017,Li-2019,Kai-2020,Brabec-2021,Menczer-2023}.
Since scaling analysis as a function of the number of GPU devices has already been reported in Ref.~\cite{Menczer-2023} here we focus only on the accessible computational power of a single node supplied with eight NVIDIA GPU devices. 
In addition, as the DMRG matrix and tensor algebra is reformulated according to non-Abelian quantum numbers, the bond dimension $D$ stands for 
the number of renormalized multiplets.
When the corresponding $U(1)$ bond dimension is also indicated it is denoted explicitly as $D_{U(1)}^{\rm eff}$. 

\subsection{Boosting the effective performance via non-Abelian symmetries}

\begin{figure}
    \centering
    \includegraphics[width=0.48\textwidth]{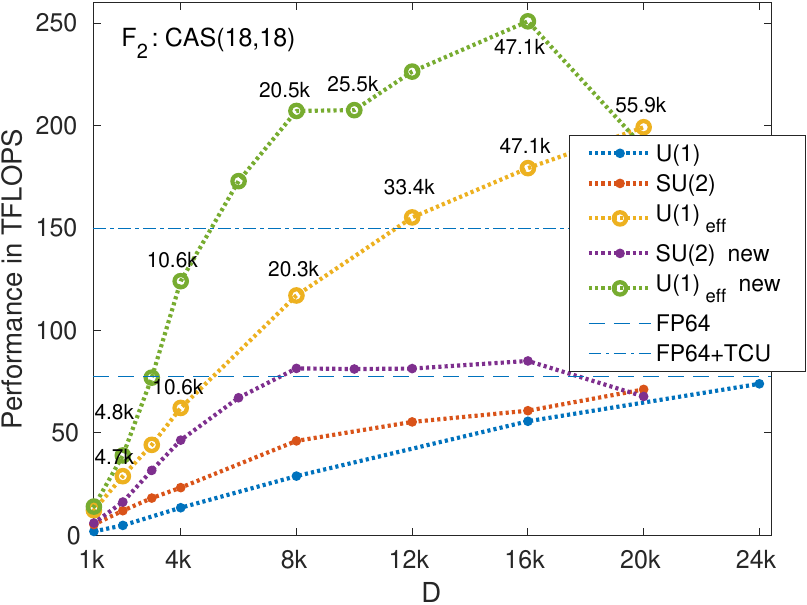}
    \caption{Benchmark results obtained via the SU(2) spin adapted hybrid CPU-multiGPU DMRG for the F$_2$ molecule for a CAS(18,18) orbital space. Calculations have been performed on a dual
    AMD EPYC 7702 CPUs with 2 × 64 cores compiled with eight NVIDIA A100-SXM4-40GB devices.
    The estimated FP64 theoretical upper bound for eight NVIDIA A100-PCIE-40GB GPU devices is shown by the dashed line while the same but also including highly specialized tensor core units (TCUs) by the dashed-dotted line. 
    The blue and red curves show the performance measured in TFLOPS for parallelization model presented in Ref.[~\cite{Menczer-2023}] using $U(1)$ and $SU(2)$ symmetries, respectively.
    Results obtained via our new parallelization model, Hash Driven Semi-Dynamic scheduling, NVIDIA, is shown by dark purple color.
    The lower bound on the effective performance in terms of $U(1)$ bond dimension after rescaling are shown by the orange and green curves.
    Numbers next to the data points show the measured largest $D_{U(1)}$ bond dimension values corresponding to $SU(2)$ bond dimension, $D$.
    }
    \label{fig:su2_f2}
\end{figure}

From technical point of view, utilization of non-Abelian symmetries leads to lower memory demands for a given accuracy threshold, denser sector decomposition of the operators and increased number of tasks, which are advantageous for massive parallelization. However, the generation of the independent tasks to be performed for the underlying algebra requires more delicate mathematical framework as discussed in Sec.~\ref{sec:theory}. 
In Fig.~\ref{fig:su2_f2} we summarize our benchmark results obtained via the $SU(2)$ spin adapted hybrid CPU-multiGPU DMRG for the F$_2$ molecule for a CAS(18,18) orbital space.
Calculations have been performed on dual AMD EPYC 7702 CPUs with 2 × 64 cores complemented with eight NVIDIA A100-SXM4-40GB devices. 
The estimated FP64 theoretical upper bound 
for eight NVIDIA A100-PCIE-40GB GPU devices~\cite{nvidia}
is shown by the dashed line while the same but also including highly specialized tensor core units (TCUs) by the dashed-dotted line. 
The blue and red curves show the performance measured in TFLOPS using $U(1)$ and $SU(2)$ symmetries
via implementation presented in Ref.~\cite{Menczer-2023}, respectively. 
These correspond to the largest performance values measured via the Davidson procedure, i.e., the sum of all the computational complexity (CC) given in TFLOP to perform the sequence of matrix and tensor operations according to Eq.~(\ref{eq:hammultpsi}) divided by the elapsed time measured in seconds. Therefore, this is an average across the most performant segment of the computation, thus peak performance can be expected to be even higher. 
It is remarkable that higher performance is measured for a given $D$ value in case of $SU(2)$ and it reaches the FP64 limit slightly faster as a function of $D$ than the $U(1)$ counterpart. 
Therefore, no additional overhead appears for matrix and tensor operations in case of the non-Abelian version. 

As the spin of the renormalized basis states is known, the corresponding effective $U(1)$ bond dimension, $D_{U(1)}^{\rm eff}$ can be calculated and monitored. In addition, having the performance measurements via the $U(1)$ implementation in hand, the ratio between the performance for a given $D$ and the corresponding $D_{U(1)}$ can be estimated. 
In Table~\ref{tab:perf} the matching data sets are collected to provide more insights
on these quantities.
\begin{table}[t]
  \centering
\begin{tabular}{l|c|c||c|c||c||r}
\hline
 \hline
$D$ & CC$_{\rm max}$ & $D_{U(1)}^{\rm eff}$ & $D_{U(1)}$ & CC$_{\rm max}$ & $R_{\rm CC}$ & $R_D$\\  
    & in GFLOP      &                      &            & in GFLOP      &       &      \\    
\hline
1024 &    680 &  2361 &  2048 &    1459 & 2.15 & 2.30 \\
2048 &   4196 &  4910 &  5120 &  20656  & 4.92 & 2.40 \\
4096 &  27586 & 10928 & 10240 &  160912 & 5.83 & 2.67 \\
8192 & 193949 & 20797 & 20480 & 1166565 & 6.01 & 2.54 \\  
 \hline
 \hline
\end{tabular}
\caption{Collected data sets to provide an estimate on the ratio, $R_{\rm CC}$, between the largest performance measured for a given $D$ and $D_{U(1)}$, and the ratio, $R_D$, between $D$ and $D_{U(1)}^{\rm eff}$. Here, CC$_{\rm max}$ stands for the largest total computational complexity measured in FLOP for a given Davidson step.}
\label{tab:perf}
\end{table}
The ratio $R_{\rm CC}$ could be used to rescale the measured $SU(2)$ performance to predict the effective performance in case of a $U(1)$ implementation only to reach the same accuracy.   
This, however, cannot be determined for large $D$ values. Since the ratio between
$D$ and $D_{U(1)}^{\rm eff}$, $R_D$, is always smaller by at least a factor of two than $R_{\rm CC}$ we could use it to provide a lower bound on the effective performance.
After rescaling the measured $SU(2)$ performance with $R_D$ 
one arrives to the curve shown by the yellow color. Numbers next to the data points indicate the corresponding largest $D_{U(1)}^{\rm eff}$ bond dimension values. Thus, this curve is an estimate of the lower bound of the effective performance in terms of $U(1)$ bond dimension, i.e., what $U(1)$ bond dimension should be used to achieve the same accuracy and what performance would such calculation correspond to.
It is worth mentioning, that for $D=24576$, i.e., for $D_{U(1)}=65535$ the exact solution, the so-called full-CI limit, could be recovered. This is basically the current limit of exact diagonalization on HPC infrastructures. Such very large $D_{U(1)}=2^{16}$ value is also the largest one utilized via Google's tensor product units (TPUs) for a non-interacting spinless fermion model \cite{Ganahl-2023}.
In general, we conclude that our framework presented for $U(1)$ symmetry in Ref.~\cite{Menczer-2023} can be utilized equally for non-Abelian symmetries, like $SU(2)$ spin symmetry, without introducing additional overhead.

\subsection{Boosting performance via new algorithmic model for parallelization}

In Ref.~\cite{Menczer-2023} we have already shown the monotonic increase in the performance as a function of the bond dimension for all selected systems, but the actual speedup utilizing GPU acceleration and its dependence on the number of GPU devices are highly system dependent (see Fig.~7 in Ref.~\cite{Menczer-2023}. 
Since the same accuracy can be reached with significantly lower $D$ values in case of $SU(2)$ symmetry it is advantageous to improve the performance for the smaller $D$ regime as well.
Although, the implementation reported in Ref.~\cite{Menczer-2023} works equally well for $U(1)$ and $SU(2)$ symmetries as discussed in the previous section (see Fig.~\ref{fig:su2_f2}) the major speedup was measured for relatively larger bond dimensions. Therefore, we have designed and developed a new mathematical model presented in Sec.~\ref{sec:methods} as a flexible framework for various types and stages of parallelisation, including MPI (multiple nodes), CUDA (multiple GPUs) and classical threading (multiple CPUs). The concurrent utilization of all physical CPU cores and NVIDIA accelerators residing within a local node is made possible by our in-house designed C++ based threading, lock-free interthread communication protocols and custom virtual device-memory mapping models.

Here we demonstrate the drastic increase in performance utilizing our novel solution 
%presented in Sec~\ref{sec:methods} via the large scale $SU(2)$ spin adapted DMRG simulation 
for the F$_2$ molecule in the CAS(18,18) orbital space.
The measured performance as a function of $D$ is shown in 
Fig.~\ref{fig:su2_f2} by the dark purple curve. The performance in most of the cases doubled at least, leading to a factor of two to three further reduction in computational time. For $D\ge 8192$, even together with the IO overhead, the measured performance is above the FP64 theoretical limit, indicating that NVIDIA TENSOR cores can also be utilized by our new parallelization model (alongside the conventional FP64 units).
Most remarkably, however, another main result is the dramatic increase in performance for smaller bond dimension values as is apparent in Fig.~\ref{fig:su2_f2}. In fact, the FP64 limit is almost reached by $D=6144$ corresponding to $D_U(1)=15755$.
We also confirmed that the measured largest computational complexity via the new
version is withing two percent agreement with data collected in Table~\ref{tab:perf}.
After resealing via $R_D$, the lower bound of the effective performance in terms of $U(1)$ symmetries, shown by the green curve, is already above $250$ TFLOPS while the actual value using $R_{\rm CC}$ could be close to $500$ TFLOPS.
This demonstrates again the tremendous increase in performance utilizing non-Abelian symmetries via our new model for parallelization.

For $D=20480$, however, a breakdown in performance is observed. This is due to the fact that for such large $D$ value the 40GB of VRAM for a single GPU device is not enough to store a meaningful amount of data, resulting in poor cache performance, which ultimately will lead to much higher IO overhead. The inability to keep data in cache even at lower levels of data abstraction as explained in Sec.~\ref{sec:methods} will likely cripple performance.  
Recall that in our new model we use the GPU memory more aggressively to boost the performance while for very large $D$ values when GPU memory is not sufficient a higher price in performance is paid.  
To simulate such scenario we have performed tests by limiting the available GPU memory to  5, 10, 15, 20, 25 GB and confirmed the systematic decrease in the number of H2D IO calls and the dramatic increase in performance.
Fortunately,
this unfavorable loss in performance can be improved or even omitted utilizing the NVIDIA fast D2D connection allowing the algorithm to treat the memory of the eight GPU devices as a single big memory unit (in the current case corresponding to 320 GB). 
Implementation details of such highly sophisticated kernel keeping track of all data segments already available in VRAM of the individual devices or still required to be transferred from the host is under development and will be part of a subsequent work.  

\begin{figure}
    \centering
    \includegraphics[width=0.48\textwidth]{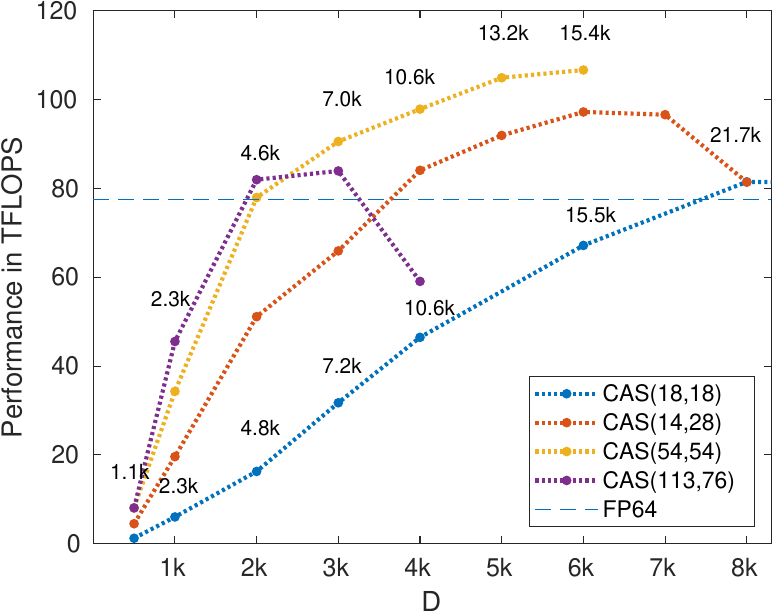}
    \caption{Benchmark results obtained via the SU(2) spin adapted hybrid CPU-multiGPU DMRG for the F$_2$ N$_2$ and FeMoco molecular systems for CAS(18,18), CAS(14,28), CAS(54,54) and CAS(113,76) orbital spaces.
    Calculations have been performed on a dual
    AMD EPYC 7702 CPUs with 2 × 64 cores compiled with eight NVIDIA A100-SXM4-40GB devices.
    The estimated FP64 theoretical upper bound for eight NVIDIA A100-PCIE-40GB GPU devices is shown by the dashed line.
    %while the same limit, now including NVIDIA TENSOR cores, is represented by the dashed-dotted line. 
    %
    }
    \label{fig:su2}
\end{figure}

In order to demonstrate that our non-Abelian symmetry related algorithmic solutions and the new technical developments are not limited to small CAS spaces in Fig~\ref{fig:su2} we present
benchmark results for much larger CAS spaces for the nitrogen dimer and for the FeMoco molecular cluster.
It is apparent that increasing the number of orbitals a much faster increase in performance can be obtained as a function of bond dimension. This is due mainly to the new SBGEMMs
protocol discussed in Sec.~\ref{sec:methods} since for each sector decomposed algebra the computational complexity increases with $N^2$. 
For the FeMoco molecular cluster for the CAS(54,54) orbital space~\cite{Reiher-2017} the measured maximum performance is around 108 TFLOPS, being again well above the FP64 limit.
For CAS(113,76) orbital space~\cite{Li-2019} similar results are obtained, however, the limitation of the 40GB VRAM size appears for smaller $D$ values.
Therefore, the breakdown in performance is experienced already for $D=4096$, which can be avoided only via the utilization of the fast D2D communication. 
Here we remark that
the performance and required VRAM heavily depend on the number of the quantum number sector based tasks and on the system size, $N$. The effect of such highly non-trivial function is clearly reflected by the different curves shown in Fig.~\ref{fig:su2}.

\subsection{Diagonalization time including IO overhead versus bond dimension}

\begin{figure}
    \centering
    \includegraphics[width=0.48\textwidth]{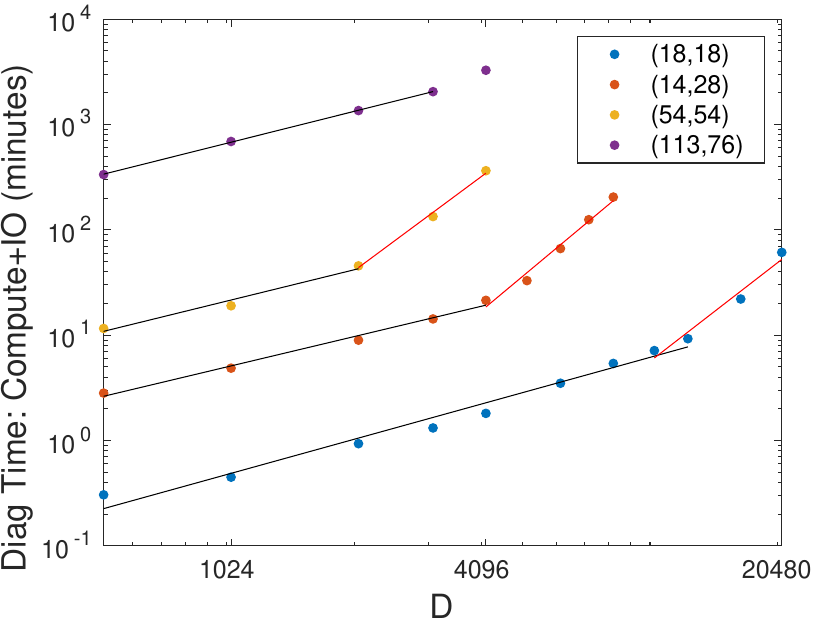}
    \caption{
    Total time of seven DMRG sweeps for the eight GPU accelerated diagonalization procedure measured in minutes including IO overhead for the F$_2$ CAS(18,18), N$_2$ CAS(14,28), FeMoco CAS(54,54) and CAS(113,76) as a function of DMRG  bond dimension. 
    The solid lines are results of ﬁrst-order polynomial ﬁts on selected data sets corresponding to measured performance up to the FP64 limit (black) and for performance above the FP64 limit (red). The fitted exponents are summarized in Tab.~\ref{tab:exp}.  
    }
    \label{fig:su2time}
\end{figure}

Completing the performance analysis as a function of the bond dimension, here we repeat the same, but for the computational time spent on the diagonalization of the effective Hamiltonian (see also Ref.~\cite{Menczer-2023}).  
The scaling of the total time spent on the eight GPU accelerated diagonalization of the effective Hamiltonian  
including H2D and D2H IO measured for seven DMRG sweeps is presented in Fig.~\ref{fig:su2time} on a double logarithmic scale as a function of DMRG bond dimension for the F$_2$, N$_2$ and FeMoco cluster for CAS(18,18), CAS(14,28), CAS(54,54) CAS(113,76) orbital spaces. The solid lines are results of ﬁrst-order polynomial ﬁts on selected data sets corresponding to calculations with performance up to the FP64 limit (black) and for $D$ values when saturation in performance is reached together with memory clearing on the GPU devices and redundant H2D IO steps (red). In the following, the two kind of exponents are denoted by $\gamma_1$ and $\gamma_2$. 

For F$_2$ the exponent is estimated to be $\gamma_1=1.11$ when no clearing step is required, while it increases to $\gamma_2=3.1$ when
saturation is reached and clearing together with redundant H2D IO protocol is utilized. 
For the larger orbital space, i.e. for N$_2$ , the linear scaling is again recovered with exponent $\gamma_1=0.96$ for $D\leq4096$ while $\gamma_2=3.3$ is obtained for larger $D$ values due to the saturation in performance. We remark that with $D=8192$ the ground state reference energy obtained by a 6th order Coupled Cluster calculation (CCSDTQPH), reported in Ref.~\cite{Chan-2004b} is also reproduced with 6 digit accuracy.
%-109.28217237847159

For the FeMoco cluster for CAS(54,54) $\gamma_1=0.98$ for $D\leq 2048$ while again
a much larger value, $\gamma_2=2.97$, is obtained when the saturation in performance is reached for $D\ge2048$. Here we note, that
in order not to waste computational resources, the 
calculations with $D=5120$ and $6144$ were stopped after the fourth sweep as performance measurements were already completed at that iteration stage and  
the saturation of the curve in Fig~\ref{fig:su2} together with the cubic scaling already obtained for the range $2048\leq D\leq 4096$ would not be effected. 
For the FeMoco cluster for CAS(113,76) $\gamma_1=0.98$ up to $D=3072$ while again
a much larger value for $\gamma_2$ is expected for $D\ge 4096$ values. 

In Tab.~\ref{tab:exetime} the measured total diagonalization time including IO overhead are collected for various parameter sets.
Here we remark that the $SU(2)$ spin adapted variant of the dynamically extended active space (DEAS) procedure  \cite{Legeza-2003b} is not developed yet. Therefore, for providing fair comparison with our previously published $U(1)$ results the total execution time is measured between the second and ninth sweep, in order to exclude the time of the non-optimized warmup procedure. This in general would have provided some 5-10 percent overhead.  
In addition, we observed that due to the lack of the optimized warmup a factor of two to three higher number of matrix vector multiplications were utilized by the Davidson algorithm for the same convergence during the third and fourth sweeps. These drawbacks will be eliminated via the $SU(2)$ adapted DEAS and RAS procedures~\cite{Legeza-2003b,Friesecke-2022b} that are also part of our current developments.
\begin{table}[t]
  \centering
\begin{tabular}{l|c|c|r}
\hline
 \hline
 System & CAS & $\gamma_1$ & $\gamma_2$ \\
\hline
F$_2$ & (18,18) & 1.11 & 3.10\\
N$_2$ & (14,28) & 0.96 & 3.3 \\
FeMoco & (54,54) & 0.98 & 2.97 \\
FeMoco & (113,76) & 1.01 & -\\
 \hline
 \hline
\end{tabular}
\caption{Fitted exponents for the eight GPU accelerated diagonalization step for the F$_2$, N$_2$ and the FeMoco molecular systems for various DMRG parameters for data sets shown in Fig.~\ref{fig:su2time}.
Exponent $\gamma_1$ corresponds to data sets measured for performance up to the FP64 limit and $\gamma_2$ to data sets with saturation in performance above the FP64 limit.
}
\label{tab:exp}
\end{table}

\begin{table}[t]
  \centering
\begin{tabular}{l|r|r|c|r}
\hline
 \hline
 CAS & $D$ & Effective  & \#of& Diag time:\\
     &     & $D_{U(1)}$ & sweeps & Compute+IO\\
 \hline
 (18,18) & 4096 & 10932 & 7 & 1.8 minutes\\
 (18,18) & 20480& 57221 & 7 & 61 minutes$^*$\\
 (14,28) & 4096& 11226 & 7 & 21 minutes\\
 (14,28) & 8192& 22243 & 7 & 3.4 hours$^*$\\
 (54,54) & 1024 & 2256  & 7 & 19 minutes \\
 (54,54) & 2048 & 4642  & 7 & 45 minutes \\
 (54,54) & 3072 & 7061  & 7 & 2.4 hours\\
 (54,54) & 4096 & 10917  & 7 & 5.7 hours\\
 (113,76) & 512 & 1502 & 7 & 5.6 hours\\ 
 (113,76) & 1024 & 3426 & 7 & 11.3 hours\\
 (113,76) & 2048 & 6965 & 7 & 22.7 hours\\
 (113,76) & 3072 & 9058 & 7 & 34 hours\\
 (113,76) & 4096 & 12101 & 7 & 54.1 hours$^*$\\
 \hline
 \hline
\end{tabular}
\caption{Total computation time together with IO overhead for the eight GPU accelerated diagonalization step for the F$_2$, N$_2$ and the FeMoco molecular systems for the spin adapted DMRG for seven sweeps and for various bond dimension values. The largest measured effective $U(1)$ bond dimension is also displayed. $^*$ indicates calculations with
increased execution time due to breakdown in performance via our current implementation facing memory limitation of the individual GPU devices.
}
\label{tab:exetime}
\end{table}

\subsection{Renormalization procedure}

As in case of our previous parallelization scheme presented for the U(1) symmetry in Ref.~\cite{Menczer-2023}, the performance of the multiGPU variant of the renormalization procedure does not scale
optimally with the number of devices for non-Abelian symmetry as well.
This is due again to the heavy D2H and H2D IO demands as
the memory requirement to store the renormalized operators scales with $N^2$.
In fact, for small and intermediate bond dimensions the IO overhead can be larger than the computational gain when several GPU devices are utilized, except for very large bond dimension values.
Compared to the CPU-only limit, however, even a singe GPU accelerated renormalization procedure becomes a factor of two to four faster.
Therefore, in most of the calculations we have used a single GPU device during the renormalization step.
Again a much more efficient renormalization algorithm will become available by utilizing the fast NVIDIA D2D communication immediately reducing the redundant H2D IO significantly. Fortunately, due to the general structure of the parallleization model discussed in Sec.~\ref{sec:methods} a single framework is developed for the various algorithmic tasks, therefore, improvements obtained for the diagonalization procedure via fast D2D communication will become immediately available for the renormalization procedure as well.
Further details will be presented in a subsequent work. 

%------------------------------------------------

\section{Conclusion}
\label{sec:conclusion}
In this work, we have presented novel algorithmic solutions together with implementation details to extend current limits of tensor network state algorithms via non-Abelian symmetries on high performance computing infrastructure. Building on state-of-the-art hardware and software technologies these includes the following main contributions:
\begin{itemize}
\item We have presented a general algorithmic design to separate the Clebsch-Gordan layer from the MPS layer 
for non-Abelian symmetries which allows efficient execution of the underlying algebra without additional overhead. As a result, the tensor and matrix algebra developed for $U(1)$ symmetry is modified only by constant multiplication factors that can be precalculated and stored in lookup tables.  

\item We have decomposed the sequence of position dependent matrix arrays (renormalized operators) into subgroups. Our custom tailored virtual memory management ensures this data is mapped into device memory space in accordance with the order of execution. The arising high spatial data locality is exploited through the use of specific sequences of strided batched matrix operations. A significantly higher level of SIMD parallelisation has been reached for the overall multiplication of the entire matrix array, since the matrices residing in these subgroups can be numerous.

\item We have introduced a new decentralized threading model, where newly launched threads put no strain on their pre-existing brethren, as all overhead of parallelism is self-contained. In our newly designed system all threads are self scheduled and make decisions autonomously. Threads are guaranteed to be able to work independently and are only loosely connected via a globally visible lock-free construct we call the Contract Book. Therefore, the computational burden of parallelization remains marginal and evenly distributed among workers, even at high thread counts.

\item A new hash based semi-dynamic scheduling protocol has been developed, enabling our model to dynamically scale between highest task throughput and lowest task group execution time. In the initial phase of the computation we can reap the benefits of having statically scheduled disjoint sets of independent jobs, while still having the late-stage option to iron out scheduling imperfections and make runtime optimizations.

\item In order to reduce IO operations dramatically, an adaptive buffering technique is used to dynamically match the level of data abstraction to system resources. A computation that normally would have failed due to insufficient memory will now become runnable by the frequent deregistration of data at lower levels of abstraction. This reduces the effective level at which cache operates. The inverse happens when available memory is abundant. Deregistrations will occur at a higher level, leading to more efficient caching. Ultimately, memory demand at a given time will be mapped to available memory at a given time. For every point in time, an elevation in memory usage will yield higher efficiency in caching, while drops in memory demand will cost us in performance. In our adaptive model the actual memory demand will match global device memory (VRAM) at all times. Instead, it is the cache performance that is variable.

\item 
We have performed benchmark calculations on Hilbert space dimensions up to $2.88\times10^{36}$ obtained via the large-scale $SU(2)$ spin adapted hybrid CPU-multiGPU density matrix renormalization group method for the F$_2$, N$_2$ and FeMoco strongly correlated molecular systems. 
These demonstrate the utilization of the highly specialized NVIDIA tensor core units (TCUs) leading to a performance around 110 TFLOPS on a single node supplied with eight NVIDIA A100 devices. The effective performance to reach the same accuracy with a $U(1)$ implementation only was estimated to be in the range of 250-500 TFLOPS.

\item We have analyzed the total computation time including IO overhead for the diagonalization procedure as a function of DMRG bond dimension and found two sets of exponents determining the overall scaling. 
For bond dimension values where the obtained performance is not exceeding the FP64 saturation limit the exponent is one leading to linear scaling. On the other hand, when a saturation or even a breakdown in performance is measured a cubic dependence is found as a function of bond dimension.  
Note that the former behavior has already been reported in Ref.~\cite{Menczer-2023} for an implementation including $U(1)$ symmetry only, while the latter regime could have been accessed only via non-Abelian symmetry, due to the big reduction in memory and the tremendous increase in computational complexity. 

\item Our new parallelization model has not only lead to significant increase in maximum performance, but also boosts the performance for small and intermediate bond dimensions. This can make the non-Abelian variant of the hybrid CPU-multiGPU DMRG to become a routinely applied daily method.
 
\end{itemize}
    
Our developments can be further improved by utilization of NVIDIA fast D2D communication allowing us to treat the total memory of the individual devices as a single unit, and thus to reduce IO overhead significantly.
In addition, combination of the hybrid CPU-multiGPU solution with our message passing inter-face (MPI) based code to achieve multiNode-multiGPU version of the algorithm will raise its performance to the petascale regime.
These developments are part of our current works and details will be published in a subsequent contribution.

%------------------------------------------------

\section*{Acknowledgement}
The authors acknowledge useful discussions with Tam\'as Kozsik, Mikl\'os Werner and Jeff Hammond.
This research was supported 
by the Hungarian National Research, Development and Innovation Office (NKFIH) through Grant Nos.~K134983 and TKP2021-NVA-04
by the Quantum Information National Laboratory of Hungary, 
and by the Hans Fischer Senior Fellowship programme funded by the Technical University
of Munich – Institute for Advanced Study. 
\"O.L. has also been supported
by the Center for Scalable and Predictive methods
for Excitation and Correlated phenomena (SPEC),
funded as part of the Computational Chemical Sciences Program by the U.S. Department of Energy
(DOE), Office of Science, Office of Basic Energy Sciences, Division of Chemical Sciences, Geosciences, and Biosciences at Pacific Northwest National Laboratory.
We thank computational support from 
the Eötvös Loránd University and
the national supercomputer HPE Apollo Hawk at the High Performance Computing Center Stuttgart (HLRS) under the grant number MPTNS/44246.

\section*{Appendix A}
\label{sec:appendix}

In this section we present a possible derivation of the correction factor, $\Tilde{C}$, that appears in the matrix and tensor algebra for SU(2) non-Abelian symmetry relying on a Matlab implementation of the Wigner-9j and Wigner-6j formalism~\cite{Varshoalovich-1988,libsu2,Weisstein-0000}, i.e., 
\begin{equation}
\begin{aligned}
    \tilde{C} = & \sqrt{(2j_1^\prime+1)(2j_2^\prime+1)(2j+1)(2k+1)}\times \\
    & W_{\rm 9j}(j_1,j_2,j,k_1,k_2,k,j_1^\prime,j_2^\prime,j^\prime) 
\end{aligned}
%\label{eq:aa}
\end{equation}
where
\begin{equation}
\begin{aligned}
W_{\rm 9j}(j_1,j_2,j,k_1,k_2,k,j_1^\prime,j_2^\prime,j^\prime) =\\ 
\sum_{x=x_{\rm min}}^{x_{\rm max}}(-1)^{2x}(2x+1)\cdot
W_{\rm 6j}(j_1,j_2,j,k,j^\prime,x)\times \\
W_{\rm 6j}(k_1,k_2,k,j_2,x,j_2^\prime)\cdot 
W_{\rm 6j}(j_1^\prime,j_2^\prime,j^\prime,x,j_1,k_1),
\end{aligned}
\label{eq:w9}
\end{equation}
with
\begin{equation}
\begin{aligned}
& x_{\rm min} = \max(j_1-j^\prime,j_2-k,k_1-j_2^\prime),\\
& x_{\rm max} = \min(j_1+j^\prime,j_2+k,k_1+j_2^\prime),
\end{aligned}
\end{equation}
\begin{equation}
\begin{aligned}
& |j_1-j_2|\leq j \leq j_1+j_2,\,\,
& |k_1-k_2|\leq k \leq k_1+k_2,\\
& |j_1^\prime-j_2^\prime|\leq j^\prime \leq j_1^\prime+j_2^\prime,\,\,
& |j_1-k_1|\leq j_1^\prime \leq j_1+k_1,\\
&|j_2-k_2|\leq j_2^\prime \leq j_2+k_2,\,\,
&|j-k|\leq j^\prime \leq j+k,
\end{aligned}
\end{equation}
and
\begin{equation}
\begin{aligned}
& W_{6j}(a,b,c,d,e,f) = 
     \Tilde{g}(a,b,c) \Tilde{g}(c,d,e) 
     \Tilde{g}(a,e,f) \Tilde{g}(b,d,f)\times \\
    & \sum_{n=n_{\rm min}}^{n_{\rm max}} \bigl( (-1)^n(n+1)!\bigr)/\bigl( 
    (n-\Tilde{n}_2)!\cdot(n-\Tilde{n}_3)!\cdot 
    (n-\Tilde{n}_4)!\times \\
    & (n-\Tilde{n}_5)!\cdot  
      (\Tilde{n}_6-n)!\cdot(\Tilde{n}_7-n)!\cdot
      (\Tilde{n}_8-n)! \bigr),
\end{aligned}
\label{eq:w6}
\end{equation}
with
\begin{equation}
\begin{aligned}
& \Tilde{n}_1 = 0, \,\, 
&  \Tilde{n}_2 = a+b+c,  \\
& \Tilde{n}_3 = c+d+e,  \,\,
&  \Tilde{n}_4 = a+e+f,  \\
& \Tilde{n}_5 = b+d+f,  \,\,
&  \Tilde{n}_6 = a+b+d+e,  \\
&  \Tilde{n}_7 = a+c+d+f,  \,\,
&  \Tilde{n}_8 = b+c+e+f,  \\
& n_{\rm min} = \max_{i=1\ldots 5}\Tilde{n}_i, \,\, 
&  n_{\rm max} = \min_{i=6\ldots 8}\Tilde{n}_i\\
& |a-b|\leq c \leq a+b, \,\,
&  |c-d|\leq e \leq c+d, \\
& |a-e|\leq f \leq a+e, \,\, 
&  |b-d|\leq f \leq b+d, \\
& {\rm mod}(a+b+c,1)=0, \,\,
&  {\rm mod}(c+d+e,1)=0,\\
& {\rm mod}(a+e+f,1)=0, \,\,
&  {\rm mod}(b+d+f,1)=0,
\end{aligned}
\end{equation}
and
\begin{equation}
\begin{aligned}
\Tilde{g}(a,b,c) = & \sqrt{(a+b-c)! (a-b+c)! (-a+b+c)!}/\\
& \sqrt{(a+b+c+1)!}.
\label{eq:del}
\end{aligned}
\end{equation}
Here we remark that derivation of $W_{9j}$ and $W_{6j}$ is also possible via directly from the Clebsch-Gordan coefficients\cite{Messiah-1962,Varshoalovich-1988,Werner-2020su2}.
%----------------------------------------------------------------------------------------
%	 REFERENCES
%----------------------------------------------------------------------------------------

\bibliographystyle{unsrt}
\bibliography{main}{}

\begin{thebibliography}{10}

\bibitem{Xu-2023-herculean}
Xiaosi Xu, Simon Benjamin, Jinzhao Sun, Xiao Yuan, and Pan Zhang.
\newblock A herculean task: Classical simulation of quantum computers, 2023.

\bibitem{Nassif-2022}
Nevine Nassif, Ashley~O. Munch, Carleton~L. Molnar, Gerald Pasdast,
  Sitaraman~V. Lyer, Zibing Yang, Oscar Mendoza, Mark Huddart, Srikrishnan
  Venkataraman, Sireesha Kandula, Rafi Marom, Alexandra~M. Kern, Bill Bowhill,
  David~R. Mulvihill, Srikanth Nimmagadda, Varma Kalidindi, Jonathan Krause,
  Mohammad~M. Haq, Roopali Sharma, and Kevin Duda.
\newblock Sapphire rapids: The next-generation intel xeon scalable processor.
\newblock In {\em 2022 IEEE International Solid- State Circuits Conference
  (ISSCC)}, volume~65, pages 44--46, 2022.

\bibitem{Burd-2022}
Thomas Burd, Wilson Li, James Pistole, Srividhya Venkataraman, Michael McCabe,
  Timothy Johnson, James Vinh, Thomas Yiu, Mark Wasio, Hon-Hin Wong, Daryl
  Lieu, Jonathan White, Benjamin Munger, Joshua Lindner, Javin Olson, Steven
  Bakke, Jeshuah Sniderman, Carson Henrion, Russell Schreiber, Eric Busta,
  Brett Johnson, Tim Jackson, Aron Miller, Ryan Miller, Matthew Pickett, Aaron
  Horiuchi, Josef Dvorak, Sabeesh Balagangadharan, Sajeesh Ammikkallingal, and
  Pankaj Kumar.
\newblock Zen3: The amd 2nd-generation 7nm x86-64 microprocessor core.
\newblock In {\em 2022 IEEE International Solid- State Circuits Conference
  (ISSCC)}, volume~65, pages 1--3, 2022.

\bibitem{Munger-2023}
Benjamin Munger, Kathy Wilcox, Jeshuah Sniderman, Chuck Tung, Brett Johnson,
  Russell Schreiber, Carson Henrion, Kevin Gillespie, Tom Burd, Harry Fair,
  David Johnson, Jonathan White, Scott McLelland, Steven Bakke, Javin Olson,
  Ryan McCracken, Matthew Pickett, Aaron Horiuchi, Hien Nguyen, and Tim~H
  Jackson.
\newblock “zen 4”: The amd 5nm 5.7ghz x86-64 microprocessor core.
\newblock In {\em 2023 IEEE International Solid- State Circuits Conference
  (ISSCC)}, pages 38--39, 2023.

\bibitem{Elster-2022}
Anne~C. Elster and Tor~A. Haugdahl.
\newblock Nvidia hopper gpu and grace cpu highlights.
\newblock {\em Computing in Science \& Engineering}, 24(2):95--100, 2022.

\bibitem{Svedin-2021}
Martin Svedin, Steven W.~D. Chien, Gibson Chikafa, Niclas Jansson, and Artur
  Podobas.
\newblock Benchmarking the nvidia gpu lineage: From early k80 to modern a100
  with asynchronous memory transfers, 2021.

\bibitem{Jouppi-2017}
N.~P. Jouppi, C.~Young, N.~Patil, G.~Agrawal D.~Patterson, R.~Bajwa, S.~Bates,
  S.~Bhatia, N.~Boden, and et~al. A.~Borchers.
\newblock Proceedings of the 44th annual international symposium on computer
  architecture, isca ’17.
\newblock page~1, 2017.

\bibitem{Jouppi-2020}
Norman~P. N.~P.~Jouppi, Doe~Hyun Yoon, George Kurian, Sheng Li, Nishant Patil,
  James Laudon, Cliff Young, and David Patterson.
\newblock A domain-specific supercomputer for training deep neural networks.
\newblock {\em Commun. ACM}, 63(7):67–78, jun 2020.

\bibitem{Feynman-1949a}
R.~P. Feynman.
\newblock The theory of positrons.
\newblock {\em Phys. Rev.}, 76:749--759, Sep 1949.

\bibitem{anzt2023high}
H.~Anzt, A.~Bienz, P.~Luszczek, and M.~Baboulin.
\newblock {\em High Performance Computing. ISC High Performance 2022
  International Workshops: Hamburg, Germany, May 29 -- June 2, 2022, Revised
  Selected Papers}.
\newblock Lecture Notes in Computer Science. Springer International Publishing,
  2023.

\bibitem{Hager-2004}
G.~Hager, E.~Jeckelmann, H.~Fehske, and G.~Wellein.
\newblock Parallelization strategies for density matrix renormalization group
  algorithms on shared-memory systems.
\newblock {\em Journal of Computational Physics}, 194(2):795--808, mar 2004.

\bibitem{Stoudenmire-2013}
E.~M. Stoudenmire and Steven~R. White.
\newblock Real-space parallel density matrix renormalization group.
\newblock {\em Physical Review B}, 87(15), apr 2013.

\bibitem{Nemes-2014}
Csaba Nemes, Gergely Barcza, Zolt{\'a}n Nagy, {\"O}rs Legeza, and P{\'e}ter
  {\relax Sz}olgay.
\newblock The density matrix renormalization group algorithm on kilo-processor
  architectures: Implementation and trade-offs.
\newblock {\em Computer Physics Communications}, 185(6):1570 -- 1581, 2014.

\bibitem{Secular-2020}
Paul Secular, Nikita Gourianov, Michael Lubasch, Sergey Dolgov, Stephen~R.
  Clark, and Dieter Jaksch.
\newblock Parallel time-dependent variational principle algorithm for matrix
  product states.
\newblock {\em Physical Review B}, 101(23), jun 2020.

\bibitem{Brabec-2021}
Jiri Brabec, Jan Brandejs, Karol Kowalski, Sotiris Xantheas, Örs Legeza, and
  Libor Veis.
\newblock Massively parallel quantum chemical density matrix renormalization
  group method.
\newblock {\em Journal of Computational Chemistry}, 42(8):534--544, 2021.

\bibitem{Zhai-2021}
Huanchen Zhai and Garnet Kin-Lic Chan.
\newblock Low communication high performance ab initio density matrix
  renormalization group algorithms.
\newblock {\em Journal of Chemical Physics}, 154(22):0021--9606, 2021.

\bibitem{Gray-2021}
Johnnie Gray and Stefanos Kourtis.
\newblock Hyper-optimized tensor network contraction.
\newblock {\em Quantum}, 5, 2021.

\bibitem{Lyakh-2022}
Dmitry Lyakh, Thien Nguyen, Daniel Claudino, and Eugene Dumitrescu.
\newblock Exatn: Scalable gpu-accelerated high-performance processing of
  general tensor networks at exascale.
\newblock {\em Frontiers in Applied Mathematics and Statistics}, 8, 2022.

\bibitem{Ganahl-2023}
Martin Ganahl, Jackson Beall, Markus Hauru, Adam~G.M. Lewis, Tomasz Wojno,
  Jae~Hyeon Yoo, Yijian Zou, and Guifre Vidal.
\newblock Density matrix renormalization group with tensor processing units.
\newblock {\em PRX Quantum}, 4:010317, 2023.

\bibitem{Liu-2023}
Minzhao Liu, Changhun Oh, Junyu Liu, Liang Jiang, and Yuri Alexeev.
\newblock Supercomputing tensor networks for u(1) symmetric quantum many-body
  systems, 2023.

\bibitem{Menczer-2023}
Andor Menczer and Örs Legeza.
\newblock Massively parallel tensor network state algorithms on hybrid cpu-gpu
  based architectures, 2023.

\bibitem{White-1992b}
Steven~R. White.
\newblock Density matrix formulation for quantum renormalization groups.
\newblock {\em Phys. Rev. Lett.}, 69:2863--2866, November 1992.

\bibitem{Schollwock-2005}
Ulrich Schollw\"ock.
\newblock The density-matrix renormalization group.
\newblock {\em Rev. Mod. Phys.}, 77:259--315, Apr 2005.

\bibitem{Noack-2005}
Reinhard~M. Noack.
\newblock Diagonalization- and numerical renormalization-group-based methods
  for interacting quantum systems.
\newblock In {\em {AIP} Conference Proceedings}. {AIP}, 2005.

\bibitem{Verstraete-2008}
F.~Verstraete, V.~Murg, and J.I. Cirac.
\newblock Matrix product states, projected entangled pair states, and
  variational renormalization group methods for quantum spin systems.
\newblock {\em Advances in Physics}, 57(2):143--224, 2008.

\bibitem{Legeza-2008}
{\"O}.~Legeza, R.M. Noack, J.~S\'olyom, and L.~Tincani.
\newblock Applications of quantum information in the density-matrix
  renormalization group.
\newblock In H.~Fehske, R.~Schneider, and A.~Weiße, editors, {\em
  Computational Many-Particle Physics}, volume 739 of {\em Lecture Notes in
  Physics}, pages 653--664. Springer, Berlin, Heidelberg, 2008.

\bibitem{Chan-2008}
Garnet Kin-Lic Chan, Jonathan~J. Dorando, Debashree Ghosh, Johannes Hachmann,
  Eric Neuscamman, Haitao Wang, and Takeshi Yanai.
\newblock An introduction to the density matrix renormalization group ansatz in
  quantum chemistry.
\newblock In Stephen Wilson, Peter~J. Grout, Jean Maruani, Gerardo
  Delgado-Barrio, and Piotr Piecuch, editors, {\em Frontiers in Quantum Systems
  in Chemistry and Physics}, volume~18 of {\em Progress in Theoretical
  Chemistry and Physics}. Springer, Netherlands, 2008.

\bibitem{Schollwock-2011}
Ulrich Schollw\"ock.
\newblock The density-matrix renormalization group in the age of matrix product
  states.
\newblock {\em Annals of Physics}, 326(1):96 -- 192, 2011.
\newblock January 2011 Special Issue.

\bibitem{Szalay-2015a}
{\relax Sz}il{\'a}rd {\relax Sz}alay, Max Pfeffer, Valentin Murg, Gergely
  Barcza, Frank Verstraete, Reinhold Schneider, and {\"O}rs Legeza.
\newblock Tensor product methods and entanglement optimization for ab initio
  quantum chemistry.
\newblock {\em Int. J. Quantum Chem.}, 115(19):1342--1391, 2015.

\bibitem{Orus-2014}
Rom\'an Or\'us.
\newblock A practical introduction to tensor networks: Matrix product states
  and projected entangled pair states.
\newblock {\em Annals of Physics}, 349(0):117 -- 158, 2014.

\bibitem{Khoromskaiaab-2015}
Venera Khoromskaiaab and Boris~N. Khoromskijb.
\newblock Tensor numerical methods in quantum chemistry: from hartree–fock to
  excitation energies.
\newblock {\em Phys. Chem. Chem. Phys.}, 17, 2015.

\bibitem{Baiardi-2020}
Alberto Baiardi and Markus Reiher.
\newblock The density matrix renormalization group in chemistry and molecular
  physics: Recent developments and new challenges.
\newblock {\em The Journal of Chemical Physics}, 152(4):040903, 2020.

\bibitem{Wigner-1959}
Eugene~Paul Wigner.
\newblock {\em Group Theory and Its Application to the Quantum Mechanics of
  Atomic Spectra}.
\newblock Academic Press, 1959.

\bibitem{Mcculloch-2002}
Ian~P McCulloch and Mikl{\'o}s Gul{\'a}csi.
\newblock The non-abelian density matrix renormalization group algorithm.
\newblock {\em Europhysics Letters}, 57(6):852, 2002.

\bibitem{Mcculloch-2007}
Ian~P McCulloch.
\newblock From density-matrix renormalization group to matrix product states.
\newblock {\em Journal of Statistical Mechanics: Theory and Experiment},
  2007(10):P10014, 2007.

\bibitem{Toth-2008}
AI~T{\'o}th, CP~Moca, {\"O}~Legeza, and G~Zar{\'a}nd.
\newblock Density matrix numerical renormalization group for non-abelian
  symmetries.
\newblock {\em Physical Review B}, 78(24):245109, 2008.

\bibitem{Legeza-2008manual}
O.~Legeza, C.~P. Moca, A.~I. Toth, I.~Weymann, and G.~Zarand.
\newblock Manual for the flexible dm-nrg code, 2008.

\bibitem{Moca-2012}
C.~P. Moca, A.~Alex, J.~von Delft, and G.~Zarand.
\newblock {\em Phys. Rev. B}, 86(195128), 2012.

\bibitem{Sharma-2012}
Sandeep Sharma and Garnet Kin-Lic Chan.
\newblock Spin-adapted density matrix renormalization group algorithms for
  quantum chemistry.
\newblock {\em The Journal of Chemical Physics}, 136(12):124121, 2012.

\bibitem{Weichselbaum-2012}
Andreas Weichselbaum.
\newblock Non-abelian symmetries in tensor networks: A quantum symmetry space
  approach.
\newblock {\em Annals of Physics}, 327(12):2972--3047, 2012.

\bibitem{Singh-2012}
Sukhwinder Singh and Guifre Vidal.
\newblock Tensor network states and algorithms in the presence of a global
  su(2) symmetry.
\newblock {\em Phys. Rev. B}, 86:195114, Nov 2012.

\bibitem{Wouters-2014}
Sebastian Wouters and Dimitri~Van Neck.
\newblock The density matrix renormalization group for ab initio quantum
  chemistry.
\newblock {\em The European Physical Journal D}, 68(9), sep 2014.

\bibitem{Keller-2016}
Sebastian Keller and Markus Reiher.
\newblock Spin-adapted matrix product states and operators.
\newblock {\em The Journal of Chemical Physics}, 144(13):134101, apr 2016.

\bibitem{Hubig-2018}
C.~Hubig.
\newblock {\em SciPost Physics}, 5(047), 2018.

\bibitem{Werner-2019}
Mikl{\'{o}}s~Antal Werner, C{\u{a}}t{\u{a}}lin~Pa{\c{s}}cu Moca, Örs Legeza,
  M{\'{a}}rton Kormos, and Gergely Zar{\'{a}}nd.
\newblock Spin fluctuations after quantum quenches in the haldane chain:
  Numerical validation of the semi-semiclassical theory.
\newblock {\em Physical Review B}, 100(3), jul 2019.

\bibitem{Gunst-2019}
Klaas Gunst, Frank Verstraete, and Dimitri~Van Neck.
\newblock Three-legged tree tensor networks with su(2) and molecular point
  group symmetry.
\newblock {\em Journal of Chemical Theory and Computation}, 15(5):2996--3007,
  apr 2019.

\bibitem{Weichselbaum-2020}
A.~Weichselbaum.
\newblock {\em Phys. Rev. Research}, 2(023385), 2020.

\bibitem{Werner-2020su2}
Mikl{\'o}s~Antal Werner, C{\u{a}}t{\u{a}}lin~Pa{\c{s}}cu Moca, {\"O}rs Legeza,
  and Gergely Zar{\'a}nd.
\newblock Quantum quench and charge oscillations in the su (3) hubbard model: A
  test of time evolving block decimation with general non-abelian symmetries.
\newblock {\em Physical Review B}, 102(15):155108, 2020.

\bibitem{Dukelsky-2004}
Jorge Dukelsky and Stuart Pittel.
\newblock The density matrix renormalization group for finite {F}ermi systems.
\newblock {\em Reports on Progress in Physics}, 67(4):513, 2004.

\bibitem{dmrg-budapest}
{\"O}rs Legeza, Libor Veis, and Tam\'as Mosoni.
\newblock {DMRG}-budapest, a program for condensed matter, quantum chemical,
  and nuclear shell {DMRG} calculations, 2022.

\bibitem{Xiang-1996}
T.~Xiang.
\newblock Density-matrix renormalization-group method in momentum space.
\newblock {\em Phys. Rev. B}, 53:R10445--R10448, Apr 1996.

\bibitem{White-1999}
Steven~R. White and Richard~L. Martin.
\newblock Ab initio quantum chemistry using the density matrix renormalization
  group.
\newblock {\em The Journal of Chemical Physics}, 110(9):4127--4130, 1999.

\bibitem{Knecht-2014}
Stefan Knecht, {\"O}rs Legeza, and Markus Reiher.
\newblock Communication: Four-component density matrix renormalization group.
\newblock {\em The Journal of Chemical Physics}, 140(4):041101, 2014.

\bibitem{Legeza-2015}
{\"O}.~Legeza, L.~Veis, A.~Poves, and J.~Dukelsky.
\newblock Advanced density matrix renormalization group method for nuclear
  structure calculations.
\newblock {\em Phys. Rev. C}, 92:051303, Nov 2015.

\bibitem{Legeza-2018a}
{\"O}rs Legeza and Christian Schilling.
\newblock Role of the pair potential for the saturation of generalized pauli
  constraints.
\newblock {\em Phys. Rev. A}, 97:052105, May 2018.

\bibitem{Shapir-2019}
I.~Shapir, A.~Hamo, S.~Pecker, C.~P. Moca, {\"O}.~Legeza, G.~Zarand, and
  S.~Ilani.
\newblock Imaging the electronic wigner crystal in one dimension.
\newblock {\em Science}, 364(6443):870--875, 2019.

\bibitem{Barcza-2020}
Gergely Barcza, Viktor Ivády, Tibor Szilvási, Márton Vörös, Libor Veis,
  Ádám Gali, and Örs Legeza.
\newblock Dmrg on top of plane-wave kohn-sham orbitals: case study of defected
  boron nitride.
\newblock {\em J. Chem. Theory Comput.}, 17(2):1143--1154, 2021.

\bibitem{Keller-2015}
Sebastian Keller, Michele Dolfi, Matthias Troyer, and Markus Reiher.
\newblock An efficient matrix product operator representation of the quantum
  chemical hamiltonian.
\newblock {\em The Journal of Chemical Physics}, 143(24), dec 2015.

\bibitem{Chan-2016}
Garnet Kin-Lic Chan, Anna Keselman, Naoki Nakatani, Zhendong Li, and Steven~R.
  White.
\newblock Matrix product operators, matrix product states, and ab initio
  density matrix renormalization group algorithms.
\newblock {\em The Journal of Chemical Physics}, 145(1):014102, 2016.

\bibitem{tenlibol}
{\"O}rs Legeza.
\newblock Matrix and tensor library for the density matrix renormalization
  group method, 1995--2023.

\bibitem{Cornwell-1997}
J.~F. Cornwell.
\newblock {\em Group Theory in Physics, An Introduction}.
\newblock Academic Press, 1997.

\bibitem{Mcculloch-2002a}
I.~P. McCulloch and M.~Gul\'acsi.
\newblock The non-{A}belian density matrix renormalization group algorithm.
\newblock {\em EPL (Europhysics Letters)}, 57(6):852, 2002.

\bibitem{Varshoalovich-1988}
A.~N.~Moskalev D.~A~Varshalovich and V.~K. Khersonskii.
\newblock {\em Quantum Theory of Angular Momentum}.
\newblock World Scientific, 1988.

\bibitem{libsu2}
Kobi.
\newblock {\em
  \url{http://www.mathworks.com/matlabcentral/fileexchange/20619}}, 2011.

\bibitem{Weisstein-0000}
Eric Weisstein.
\newblock Triangular inequalities.
\newblock {\em \url{http://mathworld.wolfram.com/TriangularInequalities.html}}.

\bibitem{Werner-2020su3}
Mikl{\'{o} }s~Antal Werner, C{\u{a}}t{\u{a}}lin~Pa{\c{s}}cu Moca, Örs Legeza,
  and Gergely Zar{\'{a}}nd.
\newblock Quantum quench and charge oscillations in the {SU}(3) hubbard model:
  A test of time evolving block decimation with general non-abelian symmetries.
\newblock {\em Physical Review B}, 102(15), oct 2020.

\bibitem{Chan-2004b}
Garnet Kin-Lic Chan, Mih\'aly K\'allay, and J\"urgen Gauss.
\newblock State-of-the-art density matrix renormalization group and coupled
  cluster theory studies of the nitrogen binding curve.
\newblock {\em The Journal of Chemical Physics}, 121(13):6110--6116, 2004.

\bibitem{Hoffman-2014}
Brian~M. Hoffman, Dmitriy Lukoyanov, Zhi-Yong Yang, Dennis~R. Dean, and
  Lance~C. Seefeldt.
\newblock Mechanism of nitrogen fixation by nitrogenase: The next stage.
\newblock {\em Chemical Reviews}, 114(8):4041--4062, 2014.
\newblock PMID: 24467365.

\bibitem{Reiher-2017}
Markus Reiher, Nathan Wiebe, Krysta~M. Svore, Dave Wecker, and Matthias Troyer.
\newblock Elucidating reaction mechanisms on quantum computers.
\newblock {\em Proceedings of the National Academy of Sciences},
  114(29):7555--7560, 2017.

\bibitem{Li-2019}
Zhendong Li, Junhao Li, Nikesh~S. Dattani, C.~J. Umrigar, and Garnet Kin-Lic
  Chan.
\newblock The electronic complexity of the ground-state of the femo cofactor of
  nitrogenase as relevant to quantum simulations.
\newblock {\em The Journal of Chemical Physics}, 150(2):024302, 2019.

\bibitem{Kai-2020}
Kai Guther, Robert~J. Anderson, Nick~S. Blunt, Nikolay~A. Bogdanov, Deidre
  Cleland, Nike Dattani, Werner Dobrautz, Khaldoon Ghanem, Peter Jeszenszki,
  Niklas Liebermann, Giovanni~Li Manni, Alexander~Y. Lozovoi, Hongjun Luo,
  Dongxia Ma, Florian Merz, Catherine Overy, Markus Rampp, Pradipta~Kumar
  Samanta, Lauretta~R. Schwarz, James~J. Shepherd, Simon~D. Smart, Eugenio
  Vitale, Oskar Weser, George~H. Booth, and Ali Alavi.
\newblock Neci: N-electron configuration interaction with an emphasis on
  state-of-the-art stochastic methods.
\newblock {\em The Journal of Chemical Physics}, 153(3):034107, 2020.

\bibitem{nvidia}
NVIDIA.
\newblock Nvidia a100 tensor core gpu.
\newblock {\em
  \url{https://www.nvidia.com/content/dam/en-zz/Solutions/Data-Center/a100/pdf/nvidia-a100-datasheet-us-nvidia-1758950-r4-web.pdf}},
  2021.

\bibitem{Legeza-2003b}
{\"O}.~Legeza and J.~S\'olyom.
\newblock Optimizing the density-matrix renormalization group method using
  quantum information entropy.
\newblock {\em Phys. Rev. B}, 68:195116, Nov 2003.

\bibitem{Friesecke-2022b}
Gero Friesecke, Gergely Barcza, and Legeza {\"O}rs.
\newblock Predicting the fci energy of large systems to chemical accuracy from
  restricted active space density matrix renormalization group calculations.
\newblock {\em arXiv:2209.14190}, 2022.

\bibitem{Messiah-1962}
Albert Messiah.
\newblock {\em Quantum Mechanics, Volume II}.
\newblock North Holland Publisher Company, Amsterdam, 1962.

\end{thebibliography}

\end{document}